\documentclass{autart}
\usepackage{graphics,graphicx,psfrag,array,eepic}
\usepackage{multirow}
\usepackage{hhline}
\usepackage{mathtools, cuted}
\usepackage[strict]{changepage}
\usepackage{epsfig}
\usepackage{amsmath}
\usepackage{amssymb}
\usepackage{color}
\usepackage{balance}
\usepackage{cite}
\usepackage{natbib}
\usepackage{times}
\usepackage{amsmath}
\usepackage{amssymb}
\usepackage{graphicx}
\usepackage{psfrag}
\usepackage{amsmath}
\usepackage{array}
\usepackage{times}
\usepackage{latexsym,theorem}
\usepackage{amsmath,amsfonts,amssymb,amsbsy}
\usepackage{tikz}
\usetikzlibrary{decorations.shapes,shapes.geometric,shadows,arrows,automata,positioning,calendar,mindmap,backgrounds,scopes,chains,er,3d,matrix,fit}

\newcommand{\R}{\mathbb{R}}
\newcommand{\B}{\mathbb{B}}
\newcommand{\X}{\mathbb{X}}
\newcommand{\N}{\mathbb{N}}
\newcommand{\W}{\mathbb{W}}
\newcommand{\s}[1]{\overline{#1}} 

\newcommand{\cS}{\mathcal{S}}

\newcommand{\cX}{\mathcal{X}}

\newcommand{\cC}{\mathcal{C}}
\newcommand{\cE}{\mathcal{E}}
\newcommand{\cN}{\mathcal{N}}
\newcommand{\cK}{\mathcal{K}}
\newcommand{\cM}{\mathcal{M}}
\newcommand{\cB}{\mathcal{B}}
\newcommand{\cL}{\mathcal{L}}
\newcommand{\cV}{\mathcal{V}}
\newcommand{\cF}{\mathcal{F}}
\newcommand{\cR}{\mathcal{R}}

\newcommand{\cZ}{\mathcal{Z}}
\newcommand{\cI}{\mathcal{I}}

\newcommand{\cD}{\mathcal{D}}
\newcommand{\cW}{\mathcal{W}}
\newcommand{\cY}{\mathcal{Y}}
\newcommand{\cG}{\mathcal{G}}
\newcommand{\cA}{\mathcal{A}}

\newcommand{\hausd}[2]{\operatorname{haus}(#1,#2)}

\newcommand{\conv}[1]{\operatorname{conv}({#1})}

\newcommand{\Out}[2]{\operatorname{Og}(#1,#2)}
\newcommand{\interior}{\operatorname{int}}
\newcommand{\gd}[2]{\operatorname{gd}(#1,#2)}

\newtheorem{lemma}{Lemma}
\newtheorem{definition}{Definition}
\newtheorem{theorem}{Theorem}
\newtheorem{proposition}{Proposition}
\newtheorem{corollary}{Corollary}
\newtheorem{assumption}{Assumption}
\newtheorem{remark}{Remark}
\newtheorem{example}{Example}

\newtheorem{Fact}{Fact}

\begin{document}
\begin{frontmatter}
\title{Invariance in Constrained Switching Systems} 
\thanks[footnoteinfo2]{Research supported by the French Community of Belgium and by the IAP network DYSCO.}
\thanks[footnoteinfo3]{R.J.  is a Fulbright Fellow and a FNRS Research Associate. He is  currently on sabbatical leave at UCLA, Department of Electrical Engineering, Los Angeles, USA.}
\author[First]{Nikolaos Athanasopoulos\thanksref{footnoteinfo2}}\ead{nikolaos.athanasopoulos@uclouvain.be},%
\author[Second]{Konstantinos Smpoukis}\ead{ece8196@students.ece.upatras.gr},
\author[First]{Rapha\"el M. Jungers\thanksref{footnoteinfo2},\thanksref{footnoteinfo3}}\ead{raphael.jungers@uclouvain.be}
\address[First]{ICTEAM Institute, UCLouvain, Louvain-la-Neuve, Belgium}
\address[Second]{Electrical and Computer Engineering Department, University of Patras, Greece}
\begin{abstract}
We study discrete time linear constrained switching systems with additive disturbances,
in which the switching may be on the system matrices,  the disturbance sets, the state constraint sets or a combination of the above.
In our general setting, a switching sequence is admissible if it is accepted by an automaton.
For this family of systems, stability does not necessarily imply the existence of an  invariant set. Nevertheless, it does imply the existence of an \emph{invariant multi-set}, which is a relaxation of invariance and the object of our work. 
First, we establish basic results concerning the characterization, approximation and computation of the minimal and the maximal admissible invariant multi-set. Second,  by exploiting the topological properties of the directed graph which defines the switching constraints, we propose invariant multi-set constructions with several benefits. We illustrate our results in benchmark problems in control.  
\end{abstract}
\end{frontmatter}

\section{Introduction}
Switching systems are studied extensively in the context of stability analysis
and control, see e.g. the surveys \cite{HaiLin09}, \cite{Shorten:07} and the monographs \cite{Liberzonbook},  \cite{Jungers:09}.
They provide a general and accurate modelling framework for many relevant real-world systems and processes, e.g., viral mutations \cite{Vargas:11},  congestion control  \cite{ShortWirth:06}, power electronics \cite{Switch:10}, networked control systems \cite{TijsHeemels:11} and others.
In addition, the study of complex systems, either hybrid or non-linear, often boils down to analyzing  switching systems \cite{GirPap:11}.
In many practical cases, the switching signal is not arbitrary. For example, in control applications  it is often possible to choose between a set of controllers
that achieve different objectives, resulting in  a minimum dwell time regime (\cite{DehgOng:12}). Similarly, we have constrained switching when a fault occurs in a control
loop which is not recovered instantaneously, suggesting a maximum dwell time regime (\cite{JungHeem:15}).
In constrained switching,
 the \emph{admissible} switching sequences are defined in a labelled, strongly connected directed graph, see e.g., \cite{WeissAlur:07}, \cite{Dai:12}, \cite{Nikos14}, \cite{MatRaph:15}, \cite{Dull:16}, \cite{CicGugPro:16}.
A switching sequence is admissible if it can be realized by the labels of the edges appearing in a walk of the graph.  
\begin{table*}[t] \label{table1}
\centering
\begin{tabular}{|c|c|c|}
\hline
 \textbf{System Type}&  \textbf{Minimal}  &  \textbf{Maximal}  \\ \hline
 \begin{tabular}[x]{@{}c@{}}    Time-invariant  dynamics \end{tabular} & \begin{tabular}[x]{@{}c@{}}  \cite{bertsekas:1972},  \cite{Kolmanovsky98}, \\  \cite{rakovic:kerrigan:kouramas:mayne:2004b}, \cite{Ong:06}  \end{tabular}   &  \\ \cline{1-2}
 \begin{tabular}[x]{@{}c@{}}   Difference  Inclusions  \end{tabular} &  \begin{tabular}[x]{@{}c@{}}  \cite{TechReport2005},\\ \cite{Kour:05}   \end{tabular}    &
 \begin{tabular}[x]{@{}c@{}}  \cite{Gutman87}, \\ \cite{Kolmanovsky98}, \\ \cite{blanchini:1994}, \\   \cite[Section 5.4]{BlanMi:08}    \end{tabular} \\ \cline{1-2}
 \begin{tabular}[x]{@{}c@{}}   Arbitrary Switching  \end{tabular} &  \begin{tabular}[x]{@{}c@{}}  \cite[Section 4.3]{TechReport2005}  \end{tabular}&    \\ \hline
 \begin{tabular}[x]{@{}c@{}} Constrained  Switching    \end{tabular}  &   \begin{tabular}[x]{@{}c@{}}  Theorems~\ref{theorem1},  \ref{theorem22},  Propositions~\ref{theorem2}, \ref{proposition2c}  \end{tabular} &  Theorem~\ref{theorem3} \\ \hline
\end{tabular}
\caption{Basic theoretical results on the maximal and minimal invariant multi-sets in comparison to the existing results for various types of autonomous linear dynamics.
For a detailed exposition we refer to \cite{BlanMi:08}.}
\end{table*}
\label{sec6}
\begin{table*}[t]
\centering
\begin{tabular}{|c|c|c|c|} \hline
\begin{tabular}[x]{@{}c@{}}\textbf{Graph transformations}  \end{tabular} & \begin{tabular}[x]{@{}c@{}}  \textbf{Maximal} \end{tabular} &\begin{tabular}[x]{@{}c@{}}\textbf{Minimal}
\end{tabular} &  \begin{tabular}[x]{@{}c@{}}  \textbf{Applications} \end{tabular} \\ \hline
Reduced Graph (\cite{Roozbehani:08}) & \begin{tabular}[x]{@{}c@{}}
Propositions~\ref{theorem9}, \ref{theorem13}, \ref{theorem14}   \end{tabular} &  \begin{tabular}[x]{@{}c@{}} Theorem~\ref{theorem6} \\ Propositions~\ref{theorem7}, \ref{theorem8} \end{tabular} & \begin{tabular}[x]{@{}c@{}} Minimum dwell time   \\  Maximum dwell time  \end{tabular}  \\ \hline
T-product Lift (\cite{MatRaph:15}) & Theorem~\ref{theorem99}(ii) &   Theorem~\ref{theorem99}(i) &  \begin{tabular}[x]{@{}c@{}}  Maximal invariant set \\ for linear systems\end{tabular}   \\ \hline
P-Path-Dependent Lift (\cite{LeeDu:06})  & Theorem~\ref{theorem11}(ii) & Theorem~\ref{theorem11}(i) & \begin{tabular}[x]{@{}c@{}}  Non-convex approximations \\of the minimal invariant set\end{tabular}\\ \hline
\end{tabular}
\caption{Description of our new algorithms for classical applications. The left column refers to the graph-theoretical construction that we use.}
\end{table*}
Although the stability and stabilizability analysis problems  are being addressed in the literature  \cite{WeissAlur:07}, \cite{Dai:12}, \cite{MatRaph:15}, \cite{Dull:16}, there is little work available dealing systematically with the safety analysis (\cite{blanchini:1999}, \cite{BlanMi:08}, \cite{Aubin:11}). Exceptions concern the cases dealing with Markov Jump Linear Systems (\cite{MejSalAriQue:15}) or systems under dwell-time specifications  (\cite{Blanch:10}, \cite{DehgOng:12}, \cite{Dehghan:13}, \cite{ZhangZhuangBraatz:16}, \cite{OngWangDehghan:16}).
In this article, we work with a relaxation of invariant\footnote{Throughout the paper 
by stability we mean asymptotic stability and by invariance we mean robust positive invariance \cite{blanchini:1999}, \cite{rakovic:kerrigan:kouramas:mayne:2004b}, also referred to as forward invariance \cite{Aubin:11}, \cite{GoeSanTee:12}  or  d-invariance  \cite{Kolmanovsky98}.}sets, namely, the  \emph{invariant multi-sets}.
As multi-set we refer to a collection of sets in one-to-one correspondence with the nodes of the graph that defines the admissible switching sequences.
Roughly, a multi-set is invariant when the trajectory of the system visits at each time instant a, possibly different, set which is dictated by a discrete variable. This variable keeps track of the switching signal sequence and represents a node on the switching constraints graph. We use \emph{forward and backward reachability multi-set sequences} to properly characterize and compute
invariant multi-sets.  Our contributions are threefold and concern the basic results on invariance for constrained switching systems, extensions and alternative computations of the invariant multi-sets, and  applications in well studied problems in control. In specific,

\noindent $\bullet$ Analogously to the seminal works in Table 1, we characterize the minimal and the maximal invariant multi-set. 
Moreover, we provide maximal invariant multi-set constructions and inner and outer approximations of the minimal invariant multi-set. \emph{In all cases, we provide upper bounds on the number of iterations required for convergence to the desired invariant multi-sets}.   

\noindent $\bullet$ We leverage combinatorial graph transformations from the recent literature (\cite{Roozbehani:08}, \cite{MatRaph:15}, \cite{BliTre:03}, \cite{LeeDu:06}) and propose alternative  invariant multi-set constructions that are either simpler or provide better approximations. We explicitly associate the minimal and maximal invariant multi-sets of the transformed systems to the corresponding ones of the original system.

\noindent $\bullet$
As illustrated in Table 2, we apply our framework to three well-studied benchmark problems in control. In detail, 
(i) we compute efficiently the maximal and minimal invariant multi-sets for systems under dwell-time specifications (\cite{DehgOng:12}, \cite{GirTab:10}),
(ii) we provide new non-convex approximations of the minimal invariant \emph{set} for switching systems
(iii)  we establish a method for computing  the maximal admissible invariant set for linear systems in a number of iterations proportional to the square root of the number of iterations needed by the classical approach \cite{BlanMi:08}.

\noindent {\textbf{Outline}}: In Section~\ref{section2} the setting is presented, together with the definitions of invariant multi-sets and the utilized reachability mappings.
In Section~\ref{section3} we characterize and compute the minimal invariant multi-set and its inner and outer $\epsilon$-approximations, both convex and non-convex.
In Section~\ref{section4} an equivalent procedure for computing the maximal invariant multi-set is established.
The concepts of the Reduced graph and the Reduced system  are exploited in Section~\ref{section5} and the correspondence of their invariance properties with the system under study is established.
In Section~6, the Lifted graph and the Lifted system are presented.
Applications are in Section~7, while the conclusions are drawn in Section~8. For ease of exposition, we have moved the proofs to the Appendix.
Some preliminary results in Sections~\ref{section3} and \ref{section4} are presented in~\cite{NikosKostasRaphael:16}.
\begin{remark}
The implementation of the algorithmic procedures proposed in the paper is in MATLAB, in an up-to-date desktop computer.The visualizations of the sets are done using the MPT3 Toolbox \cite{MPT3}. All polytopic operations in the numerical examples require either the vertex or the half-space description of a polytope. The removal of redundant vertices/hyperplanes in the description of the polytopes is performed using the Quick Hull algorithm \cite{barber1996quickhull}.
\end{remark}

\section{Preliminaries} \label{section2}
We write vectors $x,y$ with small letters and sets $\cS,\cX,\cV$ with capital letters in italics.
The ball of radius $\alpha$ of an arbitrary norm in $\R^n$ is denoted by $\B(\alpha)$.
The norm of a vector $x\in\R^n$ is $\| x\|$.
The distance between a vector $x\in\R^n$ and a compact set $\cS\subset\R^n$ is d$(x,\cS)=\min_{y\in\cS}\|x-y\|$ and the Hausdorff distance between two compact sets $\cS_1\subset\R^n$, $\cS_2\subset\R^n$
is $\hausd{\cS_1}{\cS_2}=\min\{ \max\limits_{x_1\in\cS_1}d(x_1,\cS_2), \max\limits_{x_2\in\cS_2}d(x_2,\cS_1)  \}$. The Minkowski sum between two sets
 $\cS_1$ and $\cS_2$ is denoted by $\cS_1\oplus\cS_2$, their set difference is $\cS_1\setminus \cS_2$, the interior of a set $\cS$ is denoted by $\interior{\cS}$ and its convex hull is $\conv{\cS}$.
 A C-set $\cS\subset\R^n$ is a convex compact set which contains the origin in its interior \cite{blanchini:1999}.
The cardinality of a set $\cV$ is denoted by $|\cV|$.
Let $\cG(\cV,\cE)$,or $\cG$, be a labeled directed graph with a set $\cV$ of nodes and a set $\cE$ of edges. A walk is a sequence $v_0e_1v_1\cdots v_k$ of vertices and edges of the graph such that for all $i\in\{1,...,k \}$ the edge $e_i$ has the source node $v_{i-1}$ and the destination node $v_i$. Given a walk from a node $s \in \cV$ to a node $d\in \cV$, we denote the sequence of the appearing labels by $\sigma(s,d)$ and the walk length by $|\sigma(s,d) |$. We denote the sequence of the nodes in the walk  by $m(s,d)$.
The distance  $\gd{i}{j}$ between two nodes $i\in\cV,j\in\cV$ is the length of a shortest path connecting $i$ to~$j$. 

\subsection{System and Assumptions}
We wish to study invariance and safety for systems whose switching sequences are constrained by a set of rules.
These rules are induced by a connected labelled directed graph.
We consider a set of matrices
$
\cA:=\{A_1,...,A_N \}\subset\R^{n\times n}
$
and a set of disturbance sets
$
\mathbb{W}=\{\cW_1,...,\cW_N \},
$
 $\cW_i\subset\R^n$. 
We consider a set of nodes
$
\cV:=\{1,2,...,M\}
$
and a set of edges
$
\cE=\{(s,d,\sigma):s\in\cV,d\in\cV,\sigma\in \{1,...,N\}   \},
$
where $s$ is the source node, $d$ is the destination node and $\sigma$ the label of the edge.
The set of outgoing edges of a node $s\in\cV$ is
$
\Out{s}{\cG}:=$ $  \{d\in\cV: (\exists \sigma\in \{1,...,N \}: (s,d,\sigma)\in\cE  )\}.
$
Finally, we consider a set of constraint sets
$
\X=\{ \cX_1,...,\cX_M\},
$
where $\cX_i\subset\R^n$, $i\in\{1,...,M \}$.
The System is
\begin{align}
x(t+1)& =A_{\sigma(t)}x(t)+w(t), \label{eq_sys1}\\
z(t+1) & \in \Out{z(t)}{\cG}, \label{eq_sys2}\\
(x(0),z(0)) & \in \R^n\times \cV, \label{eq_sys3} 
\end{align}
with $w(t)\in\cW_{\sigma(t)}$, $t\geq 0$, subject to the constraints 
\begin{align}
\sigma(t) & \in\{\sigma:  (z(t),z(t+1),\sigma)\in\cE  \}, \quad \forall t\geq 0, \label{eq_con1}\\
x(t) & \in \cX_{z(t)}, \quad \forall t\geq 0. \label{eq_con3}
\end{align}
We call \emph{nominal} the disturbance-free system, i.e., the system described by the difference equation $x(t+1)=A_{\sigma(t)}x(t)$ and \eqref{eq_sys2}, \eqref{eq_sys3}, subject to the constraints \eqref{eq_con1}--\eqref{eq_con3}.
The stability of the nominal system has been characterized by the introduction of the \emph{constrained joint spectral radius} $\check{\rho}(\cA,\cG)$ \cite{Dai:12},
a generalization of the joint spectral radius  (JSR) \cite{Jungers:09} of a matrix set $\cA\subset\R^{n\times n}$, which is in turn a generalization of the spectral radius of a matrix $A\in\R^{n\times n}$.
\begin{definition}[CJSR \cite{Dai:12}]\label{def_cjsr}
The constrained joint spectral radius (CJSR) of the nominal System is
$
\check{\rho}(\cA,\cG):=\lim\limits_{k\rightarrow \infty}\check{\rho}_k(\cA,\cG),
$
where 
$
\check{\rho}_k(\cA,\cG):=\max  \{  \|\prod\limits_{j=1}^k A_{i_j}\|^{1/k}:     
 \{i_j \}_{j\in[1,l]} $ \emph{is an admissible switching sequence}$ \}
$
is the maximum growth rate up to time $k$.
\end{definition}
It is shown \cite[Corollary 2.8]{Dai:12} that the nominal system is asymptotically stable if and only if $\check{\rho}(\cA,\cG)<1$ and asymptotic stability is equivalent to exponential stability.
\begin{assumption}[State constraints] \label{ass1}
The constraint sets $\cX_i\subset\R^n$, $i= 1,...,M$, are C-sets.
\end{assumption}
\begin{assumption}[Disturbances] \label{ass2}
The disturbance sets $\cW_i$, $i=1,...,N$, are C-sets.
\end{assumption}
\begin{assumption}[Stability]\label{ass3}
$\check{\rho}(\cA,\cG)<1$.
\end{assumption}
\begin{assumption} [Connectedness]\label{ass4}
$\cG(\cV,\cE)$ is strongly connected. 
\end{assumption}
Assumptions~\ref{ass1} and \ref{ass2} are followed by the plurality of works in the literature, see e.g., \cite{BlanMi:08} 
and they are respected by most of the real world systems. 
The assumption on convexity can be replaced in some cases to semi-algebraicity, for more details, see \cite{NikosRaphael:16}, \cite{NikosRaphael:16b}. The restriction that the constraint and disturbance sets contain the origin in their interior will be required for some of the theoretical derivations. Its alleviation is an active research topic, see e.g., \cite{Broucke:06}, \cite{BitsOla:14} for linear/arbitrary switching systems. 
Assumption~\ref{ass3} is necessary\footnote{See in Appendix B ways to verify it.
} since $\check{\rho}(\cA,\cG)>1$ excludes the existence of  non-trivial 
invariant multi-sets or safe sets.
The study of the limiting case $\check{\rho}(\cA,\cG)=1$, although interesting\footnote{Existence of invariant sets even for the case of arbitrary switching systems is undecidable (\cite{BlondTsits:00}) in this case.}, is outside the scope of this study.
Assumption~\ref{ass4} concerns the structure of the constraints in the switching signal and holds in many	 interesting cases.
\subsection{Invariant multi-sets}
We first recall the notion of an invariant set, and then generalize it to \emph{multi-sets}.
\begin{definition}[Invariance, \cite{blanchini:1999}]\label{def_invariance}
A set $\cS\subset\R^n$ is called \emph{invariant} with respect to the System \eqref{eq_sys1}-\eqref{eq_sys3}
if $x(0)\in\cS$ implies $x(t)\in\cS$, for any initial condition $z(0)\in\cV$ and any switching signal $\sigma(t), t\geq 0 $, satisfying \eqref{eq_con1}.
If additionally there is a constraint set $\cX\subset\R^n$  and $\cS\subseteq\cX$, the set $\cS$ is called \emph{admissible invariant} with respect to the System \eqref{eq_sys1}-\eqref{eq_sys2} and the constraint set.
\end{definition}
\begin{definition}[Multi-set invariance] 
The collection $\{\cS^i \}_{i\in\cV}$ is called an \emph{invariant multi-set} with respect to the System \eqref{eq_sys1}-\eqref{eq_sys3} if $x(0)\in\cS^{z(0)}$ implies $x(t)\in\cS^{z(t)}$, for all $t\geq 0$,
for any initial condition $z(0)\in\cV$ and for any switching signal $\sigma(t), t\geq 0 $, satisfying \eqref{eq_con1}.
If additionally $\cS^i\subseteq\cX_i$, for all $i\in\cV$,  the multi-set $\{\cS^i\}_{i\in\cV}$ is called an \emph{admissible invariant multi-set} with respect to the System \eqref{eq_sys1}-\eqref{eq_sys3} and the constraints $\eqref{eq_con1}, \eqref{eq_con3}$.
The admissible invariant multi-set $\{\cS_M^i\}_{i\in\cV}$ is the \emph{maximal admissible invariant multi-set} if for  any admissible invariant multi-set  $\{\cS^i\}_{i\in\cV}$ it holds that $\cS^i\subseteq\cS_M^i$, for all $i\in\cV$.
The invariant multi-set $\{\cS_m^i\}_{i\in\cV}$ is the \emph{minimal invariant multi-set} if for any invariant multi-set  $\{\cS^i\}_{i\in\cV}$ it holds that $\cS_m^i\subseteq\cS^i$, for all $i\in\cV$.
\end{definition}
\begin{definition}[Safety]
A set $\cS_{\cY}\subset\R^n$ is called \emph{safe} with respect to the System \eqref{eq_sys1}-\eqref{eq_sys3}, the constraints \eqref{eq_con1}, \eqref{eq_con3}
and with respect to  a set of nodes $\cY\subseteq\cV$ if $(x(0),z(0))\in\cS_{\cY}\times \cY$, implies $x(t)\in\cX_{z(t)}$,  for any switching signal $\sigma(t), t\geq 0 $, satisfying \eqref{eq_con1}.
The safe set $\cS^\star_{\cY}$ is called the \emph{maximal safe set} if for any other safe set  $\cS_\cY\subseteq\cX$ it holds that $\cS_\cY\subseteq\cS^\star_{\cY}$.
\end{definition}
 Assumptions~\ref{ass1}-\ref{ass4} are not sufficient for the  System \eqref{eq_sys1}-\eqref{eq_sys3} subject to the constraints \eqref{eq_con1}, \eqref{eq_con3} to possess a non-trivial invariant set, as shown in the following example.
\begin{example} We consider  the disturbance-free scalar system $x(t+1)=a_{\sigma(t)}x(t)$, with
$a_1=-2$, $a_2=0.25$ and constraint graph $\cG(\cV,\cE)$ with $\cV=\{1,2 \}$, 
$\cE=\{(1,1,2), (1,2,1), (2,1,2) \}$. The system does not admit an invariant set. Nevertheless, it admits an invariant multi-set, e.g., $\{\cS^1,\cS^2 \}=\{[-0.5,0.5],[-1,1] \}$.
\end{example}

\begin{example}\label{example1}
We consider a System \eqref{eq_sys1}--\eqref{eq_sys3}, subject to state
constraints $x(t)\in\cX$.
The switching signal may take two values $\sigma(\cdot):\N\rightarrow \{1,2 \}$. The switching constraints graph $\cG(\cV,\cE)$ is in Figure~\ref{f1}. 
For $z(0)=a$ and any $x(0)\in\R^2$ the switching sequence $\{\sigma(0),...,\sigma(6)\}=\{2,2,1,1,2,1,1\}$ is admissible  since it can be realized by the walk $(a,2,a,2,a,1,b,1,c,2,b,1,c,1,a)$, whereas the switching sequence $\{\sigma(0),\sigma(1),\sigma(2)\}=\{2,1,2\}$ is not admissible.
 \begin{figure}[ht]
\begin{center}
\begin{tikzpicture}[->,>=stealth',shorten >=1pt,auto,node distance=2.1cm, font=\small, semithick]
\tikzstyle{every state}=[fill=gray!20,draw=black,thick,text=black,,scale=0.8]
\node[state,node distance=2.5cm]         (A)  {a};
\node[state,draw=black]         (B) [ right of =A] {b};
\node[state,draw=black]         (C) [ right of =B] {c};
\path (A) edge [bend left] node {$1$} (B);
\path (A) edge [loop left] node {$2$} (A);
\path (B) edge [bend left] node {$1$} (C);
\path (C) edge [bend left] node[above] {$2$} (B);
\path (C) edge [bend left] node {$1$} (A);
\end{tikzpicture}
\end{center}
\caption{Example 1, the switching constraints graph $\cG(\cV,\cE)$.}
\label{f1}
\end{figure}
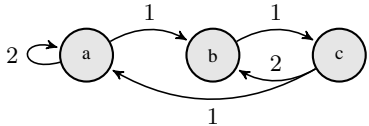
An admissible invariant multi-set $\{\cS^a,\cS^b,\cS^c\}$ for the system is depicted in Figure~\ref{f2}. An illustration of  a trajectory $x(0),...,x(6)$ is also depicted for
 the initial conditions $z(0)=a$, $x(0)\in\cS^a$,
corresponding to the switching sequence$\{\sigma(0),...,\sigma(5)\}=\{2,2,1,1,2,1,1\}$.
 \begin{figure}[ht]
\psfrag{S1}[][][0.8]{\color{black}{$\cS^a$}}
\psfrag{S2}[][][0.8]{\color{black}{$\cS^b$}}
\psfrag{S3}[][][0.8]{\color{black}{$\cS^c$}}
\psfrag{X}[][][0.8]{\color{black}{$\cX$}}
\psfrag{x(0)}[][][0.7]{\color{black}{$x(0)$}}
\psfrag{x(1)}[][][0.7]{\color{black}{$x(1)$}}
\psfrag{x(2)}[][][0.7]{\color{black}{$x(2)$}}
\psfrag{x(3)}[][][0.7]{\color{black}{$x(3)$}}
\psfrag{x(4)}[][][0.7]{\color{black}{$x(4)$}}
\psfrag{x(5)}[][][0.7]{\color{black}{$x(5)$}}
\psfrag{x(6)}[][][0.7]{\color{black}{$x(6)$}}
\psfrag{Node 1}[][][0.7]{\color{black}{Node a}}
\psfrag{Node 2}[][][0.7]{\color{black}{Node b}}
\psfrag{Node 3}[][][0.7]{\color{black}{Node c}}
\begin{center}
\includegraphics[height=4cm,width=8cm]{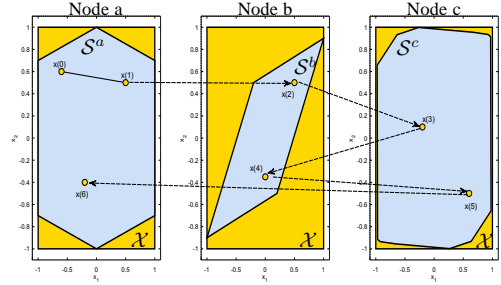}
\end{center}
\caption{Example 1, the invariant multi-set $\{\cS^a,\cS^b,\cS^3\}$ (blue),
the state constraints $\cX$ (yellow) and a trajectory for initial conditions $x(0)\in\cS^a$, $z(0)=a$.}
\label{f2}
\end{figure}
\end{example}
A natural connection of multi-set invariance to standard invariance can be made considering the unconstrained switching system,   which is associated to the original system via the so-called $\Omega$-lift and the Kronecker lift (\cite{Kozyakin:14}, \cite{Dull:16}). Although one may be tempted to work in the lifted space and apply well-established set-theoretic methods to perform the safety analysis, the respective constructions are computationally prohibitive and they do not take into account the structure of the switching constraints realized by the directed graph that defines them. Additionally, by maintaining the information of the switching constraints graph we may retrieve more refined notions related to safety, e.g., returnability (\cite{Kolmanovsky:02}) and recurrence (\cite{Teel:14}).
Another interpretation of the multi-set invariance can be made in the framework of hybrid systems \cite[Chapters 2.4, 6, 8]{GoeSanTee:12}, \cite{GoeSanTee:08}.
Also, the existence of invariant multi-sets can be associated to the  multi-norms used in \cite{MatRaph:15}.   
It is worth noting that although the sublevel sets of multi-norm Lyapunov functions constitute an invariant multi-set, approximation and exact computation of minimal/maximal/safe invariant multi-sets was not sought there.

\subsection{Reachability mappings}
We describe the standard mappings \cite{BlanMi:08}, \cite{Aubin:11} used in the paper. Consider a set of matrices $\cA\subset\R^{n\times n}$, the set of disturbance sets $\W$ and a switching sequence
 $\{\sigma_{i} \}_{i\in\{1,...,p \}}$, $\sigma_i\in\{1,...,N \}$, where $p\geq 1$.
The \emph{$p$-step forward reachability mapping} is
$\cR(\{\sigma_i \}_{i\in\{1,...,p \}}, \cS )  = (   \prod_{i=1}^p A_{\sigma_{p+1-i}}\cS     )
\oplus (    \bigoplus\limits_{j=1}^p \prod\limits_{i=1}^{p-j} A_{{\sigma_{p+1-j}}}\cW_{\sigma_{j}}  ).$
In the absence of an additive term, i.e., when $\W:=\{0\}$, we write 
$\cR_{\text{N}}\left(\{\sigma_i\}_{i\in[p]},\cS\right)  :=\left\{ \prod_{i=1}^{p}A_{\sigma_{p+1-i}}x: x\in\cS \right\}.$
Moreover, we define the \emph{`convexified'} versions of the forward mappings, i.e.,
$\cR_{\text{C}}(\{\sigma_i\}_{i\in\{1,...,p \}},\cS)  :=\conv{\cR(\{\sigma_i\}_{i\in\{1,...,p \}},\cS)}$, $\cR_{\text{CN}}(\{\sigma_i\}_{i\in\{1,...,p \}},\cS) :=\conv{\cR_{\text{N}}(\{\sigma_i\}_{i\in\{1,...,p \}},\cS)}$.
Similarly, we define the \emph{ $p$-step backward reachability mapping} as
$ \cC(\{\sigma_i\}_{i\in\{1...,p \}},\cS)  :=\{ x: \left(   \prod_{i=1}^p A_{\sigma_{p+1-i}}\{x  \}     \right)
\oplus    (    \bigoplus\limits_{j=1}^p \prod\limits_{i=1}^{p-j} A_{{\sigma_{p+1-j}}}\cW_{\sigma_{j}}             ) \in \cS
  \}$.
\begin{example} \label{example_rc}
We illustrate how the reachability mappings are applied along a switching sequence in a graph. By considering the sequence $\{ \sigma_1,\sigma_2\}$, we have
$\cR(\{ \sigma_1,\sigma_2\}, \cS)  = A_{\sigma_2}A_{\sigma_1}\cS\oplus A_{\sigma_2}\cW_{\sigma_1}\oplus\cW_{\sigma_2}$ and 
$\cC( \{\sigma_1, \sigma_2 \}, \cS  )  = \{ x: A_{\sigma_2}A_{\sigma_1}\{ x\}\oplus A_{\sigma_2}\cW_{\sigma_1}\oplus \cW_{\sigma_2} \in\cS   \}$.
\begin{figure}[h!] \label{fig3}
\begin{center}
\begin{tikzpicture}[->,>=stealth',shorten >=1pt,auto,node distance=2.2cm, font=\small, semithick]
\tikzstyle{every state}=[fill=gray!20,draw=black,thick,text=black,,scale=0.6]
\node[state,node distance=2cm]         (A)  {$\cS$};
\node[state,draw=black]         (B) [ right of =A] {};
\node[state,draw=black]         (C) [ right of =B] {$\cR(\{ \sigma_1,\sigma_2 \},\cS  )$};
\path (A) edge [bend left] node {$\sigma_1$} (B);
\path (B) edge [bend left] node {$\sigma_2$} (C);
\end{tikzpicture}
\hspace{0.5cm}
\begin{tikzpicture}[->,>=stealth',shorten >=1pt,auto,node distance=2.2cm, font=\small, semithick]
\tikzstyle{every state}=[fill=gray!20,draw=black,thick,text=black,scale=0.6]
\node[state,node distance=2cm]         (A)  {$\cC(\{ \sigma_1,\sigma_2 \},\cS  )$};
\node[state,draw=black]         (B) [ right of =A] {};
\node[state,draw=black]         (C) [ right of =B] {$\cS$};
\path (A) edge [bend left] node {$\sigma_1$} (B);
\path (B) edge [bend left] node {$\sigma_2$} (C);
\end{tikzpicture}
\end{center}
\caption{Example \ref{example_rc}, illustration  of a $2$-step forward (left) and backward  (right) reachability map.}
\label{f5}
\end{figure}
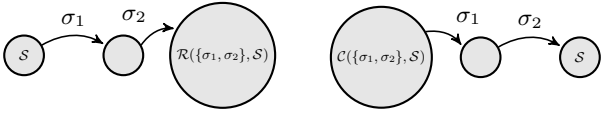
\end{example}
Using the forward and backward reachability mappings, we can verify invariance of a multi-set in a straightforward manner. Proposition~\ref{prop_altdef} follows directly from the Defininition~3.
\begin{proposition} \label{prop_altdef}
Consider a collection $\{ \cS_i\}_{i\in\cV}$ of sets and the System \eqref{eq_sys1}--\eqref{eq_sys3} subject to the constraint \eqref{eq_con1}.
The following statements are equivalent.
\begin{itemize}
\item[1.] The multi-set $\{\cS_i \}_{i\in\cV}$ is invariant with respect to the System.
\item[2.] For any edge $(i,j,\sigma)\in\cE$, it holds that $\cR(\sigma,\cS_i)\subseteq \cS_j$.
\item[3.] For any edge $(i,j,\sigma)\in\cE$, it holds that $\cC(\sigma,\cS_j)\supseteq \cS_i$.
\end{itemize}
\end{proposition}

\subsection{Stability metrics for the nominal system} \label{subsection2.5}
Assumption~\ref{ass3} implies exponential stability of the nominal system, namely the existence of two scalars $\Gamma\geq 1$ and $\rho \in (0,1)$ such that
\begin{equation} \label{eq_exprel}
 \resizebox{0.9\hsize}{!}{%
$
\|x(t) \|\leq \Gamma \rho^t \| x(0)\|,  \forall (x(0),z(0))\in \R^n\times \cV, \forall t\geq 0,
$
}
\end{equation}
 $x(\cdot)$ being any solution of the nominal system. Throughout, we utilize these stability metrics to bound the number of iterations required to compute invariant multi-sets.
We consider the sequence $\{\cN^j_l \}_{j\in\cV}$, $l\geq 0$, generated by
\begin{align}
\cN_0^j  & :=\bigcup\limits_{(s,j,\sigma)\in\cE}\cW_\sigma , \quad  j\in\cV,  \label{eq_setseq1_1} \\
\cN_{l+1}^j& := \bigcup\limits_{(s,j,\sigma)\in\cE} \cR_{\text{N}}(\sigma,\cN_l^s), \ \ \ \ j\in\cV. \label{eq_setseq1_2}
\end{align}
Equivalently to the exponential decrease of the vector norm \eqref{eq_exprel},
we can formulate the exponential contraction of the elements of the multi-set sequence
\eqref{eq_setseq1_1}, \eqref{eq_setseq1_2} in the set inclusion
\begin{equation}\label{eq_lemma1_1}
\cN_t^j\subseteq  \Gamma \rho^t\cN_0^j, \quad \forall j\in\cV, \quad \forall t\geq 0.
\end{equation}
Several methods exist for computing the scalars $\Gamma\geq 1$, $\rho\in (0,1)$ in  \eqref{eq_lemma1_1}, see, e.g., \cite{Nikos14}, \cite{MatRaph:15}, \cite{CSS:15}.  

\section{The minimal invariant multi-set} \label{section3}

In this section we characterize the minimal invariant multi-set for linear constrained switching systems.
In specific, we provide multi-set sequences for inner and outer $\epsilon$-approximations of the minimal invariant multi-set.
\subsection{Inner $\epsilon$-approximations}
We consider the sequence of multi-sets $\{\cF_l^j  \}_{j\in\cV}$, $l\geq 0$, with
\begin{align}
\cF_0^j & :=\{0\}, \quad j\in \cV  \label{eq_setseq_min1_1}, \\
\cF_{l+1}^j & :=\bigcup\limits_{(s,j,\sigma)\in\cE} \cR(\sigma,\cF_l^s), \quad j\in\cV. \label{eq_setseq_min1_2}
\end{align}
We show that this multi-set sequence converges, not necessarily in finite time\footnote{To see this, take for example  $x(t+1)=ax(t)+w(t)$, $w(t)\in[-1,1]$, for some $a\in (0,1)$.
We have $\cF_l=\bigoplus\limits_{i=0}^{l-1}(a\cF_{i}\oplus[-1,1])=[-\frac{1-a^l}{1-a},\frac{1-a^l}{1-a}]$.
Clearly, $\cS_m=\lim\limits_{l\rightarrow \infty}\cF_l=[-\frac{1}{1-a},\frac{1}{1-a}]$, however, there is no integer $k^\star\geq 0$ such that $\cF_{k^\star+1}=\cF_{k^\star}$. }, to the minimal invariant multi-set.
 Some technical observations are required first.
\begin{Fact} \label{factt1}
Consider the multi-set sequence \eqref{eq_setseq_min1_1}, \eqref{eq_setseq_min1_2}. Under Assumption~\ref{ass2}, for all $l\geq 0$ it holds
\begin{equation}\label{eq_fact1_1}
\cF_l^j=\bigcup_{i\in\{0,...,l\}}\cF_i^j, \quad \forall j\in\cV.
\end{equation}
\end{Fact}
\begin{proposition}\label{proposition1}
Consider the multi-set sequence \eqref{eq_setseq_min1_1}, \eqref{eq_setseq_min1_2}. Under Assumptions~\ref{ass2} and \ref{ass3}, there exist scalars $\rho\in (0,1)$,
 $\Gamma\geq 1$ such that  for any $l\geq 0$, it holds
\begin{equation}\label{eq_proposiition1_1}
\cF_l^j\subseteq \cF_{l+1}^j\subseteq \cF_l^j\oplus\left(\Gamma \rho^l \cW^\star \right),
\end{equation}
where $\cW^\star:=\cup_{i\in\{1,...,N \}}\cW_i$.
\end{proposition}
\begin{theorem}\label{theorem1}
Consider the multi-set sequence \eqref{eq_setseq_min1_1}, \eqref{eq_setseq_min1_2}. Under Assumptions~\ref{ass1}-\ref{ass4}, the following hold.
\begin{itemize}
\item[(i)] The sequence is convergent, i.e., there are sets $\cF_\infty^j$, $j\in\cV$, such that  $\lim\limits_{l\rightarrow \infty}\cF_l^j=\cF_{\infty}^j$.
\item[(ii)] Let
$\alpha:=\min\left\{\alpha:  \cup_{i\in \{1,...,N \}}\cW_i\subseteq \alpha \B(1)\right\}$ and a pair $(\Gamma,\rho)$ satisfying \eqref{eq_lemma1_1}.
For any $\epsilon>0$, $l\geq \left\lceil \log_{\rho} \left( \frac{\epsilon(1-\rho)}{\alpha \Gamma} \right)  \right\rceil$, it holds that
\begin{equation}  \label{eq_theorem1_1}
\cF_l^j\subseteq \cF_\infty^j\subseteq \cF_l^j\oplus\B(\epsilon),\quad \forall j\in\cV.
\end{equation}
\item[(iii)] The sequence converges to the minimal compact invariant multi-set with respect to the System \eqref{eq_sys1}-\eqref{eq_sys3} and constraint \eqref{eq_con1}, i.e.,
$\cS_m^j=\cF_\infty^j$, for all $j\in\cV$.
\end{itemize}
\end{theorem}

From Theorem~\ref{theorem1} and Proposition~\ref{proposition1}, we observe that if  $\cF_{\s{k}}^j=\cF_{\s{k}+1}^j$, for all $j\in\cV$ and an integer $\s{k}$,
then  $\cS_m^j=\cF_{\s{k}}^j$, $j\in\{1,...,M \}$.
However, since only asymptotic convergence of the multi-set sequence is guaranteed, we can utilize Theorem~\ref{theorem1}(ii) to compute \emph{$\epsilon$-inner approximations of the minimal invariant multi-set}.
 Theorem~\ref{theorem1}(ii) also provides an upper bound on the number of iterations required for any given desired accuracy $\epsilon$. The upper bound depends on the stability metrics of the System and the shapes of the disturbance sets.  Thus, it gives a new insight to the corresponding results from the literature that concern the case of arbitrary switching, e.g., \cite[Proposition 6.9]{BlanMi:08}, \cite[Section 2]{Ong:06}, where such bounds were not made explicit.

It is difficult to compute the elements of the multi-set sequence \eqref{eq_setseq_min1_1}, \eqref{eq_setseq_min1_2} as each member is a radially convex set. 
To alleviate this computational burden and in the same spirit as, e.g.,  \cite{rakovic:kerrigan:kouramas:mayne:2004b}, \cite{TechReport2005}, we turn our attention to the minimal \emph{convex} invariant multi-set. To this purpose, we consider the multi-set sequence
\begin{align}
\s{\cF}_0^j & :=\{0\}, \quad j\in\cV \label{eq_setseq_min2_1} \\
\s{\cF}_{l+1}^j & :=\bigcup\limits_{(s,j,\sigma)\in\cE}\cR_{\text{C}}(\sigma,\cF_l^s), \quad j\in\cV.  \label{eq_setseq_min2_2}
\end{align}
\begin{proposition}\label{theorem2}
Consider the multi-set sequences \eqref{eq_setseq_min2_1}, \eqref{eq_setseq_min2_2}. Under Assumptions~\ref{ass1}-\ref{ass4}, the following hold:
(i) $\conv{\cF_l^j}=\conv{\s{\cF}_l^j}$, $\quad \forall l\geq 0, \quad \forall j\in\cV$,
(ii) The multi-set sequence is convergent, there are sets $\s{\cF}_\infty^j$, $j\in\cV$, such that  $\lim\limits_{l\rightarrow \infty}\s{\cF}_l^j=\s{\cF}_{\infty}^j$,
(iii) The multi-set $\{  \conv{\s{\cF}_\infty^j}  \}_{j\in\cV}$
 is the minimal convex invariant multi-set with respect to the System \eqref{eq_sys1}-\eqref{eq_sys3} and the constraint \eqref{eq_con1},
 (iv) Let
$\alpha:=\min\left\{\alpha:  \cup_{i\in\{1,...,N \}}\cW_i\subseteq \alpha \B(1)\right\}$ and a pair $(\Gamma,\rho)$ satisfying \eqref{eq_lemma1_1}. Then,
for any $\epsilon>0$ and $l\geq \left\lceil \log_{\rho} \left( \frac{\epsilon(1-\rho)}{\alpha \Gamma} \right)  \right\rceil$ the relation 
$\conv{\s{\cF}_l^j}\subseteq \conv{\s{\cF}_\infty^j}\subseteq \conv{\s{\cF}_l^j}\oplus\B(\epsilon), j\in\cV,
$ holds.
\end{proposition}

\subsection{Invariant, outer $\epsilon$-approximations}\label{subsection3.2}
In this subsection we adapt the approach in   \cite{rakovic:kerrigan:kouramas:mayne:2004b} to the case of constrained switching systems and provide both convex and non-convex \emph{invariant} outer $\epsilon$--approximations of the minimal invariant multi-set.
We define the sets $\cN_\cap^j:=\bigcap\limits_{(i,j,\sigma)\in\cE}\cW_{\sigma}, \quad j\in\cV$.
Under Assumptions \ref{ass2}--\ref{ass4}, from \cite[Theorem 1]{Nikos14}, we have that for  any given $\lambda\in (0,1)$, there exists an integer $k\geq 1$ such that
\begin{equation}\label{eq_prop2_1}
\cN_k^j\subseteq \lambda \cN_\cap^j,  \quad  \forall j\in \cV,
\end{equation}
where the multi-set sequence $\{ \cN_l^j\}_{j\in\cV}$ is generated by   \eqref{eq_setseq1_1}, \eqref{eq_setseq1_2}.
Let $k\geq 1$, $\lambda\in [0,1)$ be such that \eqref{eq_prop2_1} holds.

\begin{theorem}\label{theorem22}
Consider the System \eqref{eq_sys1}--\eqref{eq_sys3} subject to the constraints \eqref{eq_con1} and suppose that Assumptions~2--4 hold. Let $(k,\lambda)$, $k\geq 1$, $\lambda\in (0,1)$ be such that \eqref{eq_prop2_1} holds. Consider the multi-set
$\{\cD_k^j \}_{j\in\cV}$, where
\begin{equation}\label{eq_prop2_2}
\cD_k^j:=\frac{1}{1-\lambda}\cF_{k-1}^j, \quad  j\in\cV,
\end{equation}
and $\{ \cF_l^j\}_{j\in\cV}$ is generated by
\eqref{eq_setseq_min1_1}, \eqref{eq_setseq_min1_2}. 
The following statements hold.
\begin{itemize}
\item[(i)] The multi-set $\{ \cD_k^j\}_{j\in\cV}$
is invariant with respect to the System \eqref{eq_sys1}--\eqref{eq_sys3} and the constraint \eqref{eq_con1}. 
\item[(ii)] Given a desired accuracy $\epsilon>0$ consider the pair $(k,\lambda)$ such that additional to \eqref{eq_prop2_1} they satisfy 
$\frac{\lambda}{1-\lambda}\cF_{k-1}^j\subseteq \B(\epsilon)$. Then, it holds that
\begin{equation} \label{eq_lemma2_1}
\cS_m^j\subseteq \cD_k^j\subseteq \cS_m^j\oplus\B(\epsilon), \quad \forall j\in\cV.
\end{equation}
\end{itemize}
\end{theorem}

As in the case of  inner approximations, we provide \emph{convex} outer $\epsilon$-approximations of the minimal \emph{convex} invariant multi-set, utilizing the `convexified'
versions of the forward reachability multi-set sequences $\{\s{\cF}_{l}^j \}_{j\in\cV}$  \eqref{eq_setseq_min2_1}, \eqref{eq_setseq_min2_2} and 
$\{ \s{\cN}_l^j\}_{j\in\cV}$, where
\begin{align}
\s{\cN}_0^j & =\bigcup\limits_{(s,j,\sigma)\in \cE}\cW_{\sigma},\quad j\in\cV, \label{eq_convnom1} \\
\s{\cN}_{l+1}^j & = \bigcup\limits_{(s,j,\sigma)\in\cE} \cR_{\text{CN}}(\sigma,\s{\cN}_{l}^s), \quad j\in\cV. \label{eq_convnom2}
\end{align}
Under Assumptions \ref{ass2}--\ref{ass4}, 
for any given $\lambda\in (0,1)$, there exists an integer $k\geq 1$ such that
\begin{equation}\label{eq_prop2_4}
\s{\cN}_k^j\subseteq \lambda \cN_\cap^j,  \quad  \forall j\in \cV.
\end{equation}
Let $k\geq 1$, $\lambda\in [0,1)$ such that \eqref{eq_prop2_4} holds.
 The proof of Proposition~\ref{proposition2c} is omitted since it is similar to the one of Theorem~\ref{theorem22}.
\begin{proposition} \label{proposition2c}
Consider the System \eqref{eq_sys1}--\eqref{eq_sys3} subject to the constraints \eqref{eq_con1}
and suppose that Assumptions \ref{ass2}-\ref{ass4} hold. Consider the multi-set 
$\s{\cD}_k^j:=\frac{1}{1-\lambda} \conv{\s{\cF}_{k-1}^j},  \quad j \in \cV,
$
where $\{\s{\cF}_l^j \}_{j\in\cV}$, $l\geq 0$ is generated by \eqref{eq_setseq_min2_1}, \eqref{eq_setseq_min2_2} and $k\geq 1$,  $\lambda\in (0,1)$ satisfy \eqref{eq_prop2_4}.
Given any scalar $\epsilon>0$, the relation
\begin{equation} \label{eq_th4_1}
\conv{\s{\cS}_m^j}\subseteq \s{\cD}_k^j	\subseteq \conv{\s{\cS}_m^j}\oplus\B(\epsilon), \quad \forall j\in\cV
\end{equation}
holds, for any pair $(k,\lambda)$, $k\geq 1$,  $\lambda\in (0,1)$ which satisfy (i) the relation \eqref{eq_prop2_4}, (ii) $\frac{\lambda}{1-\lambda}\s{\cF}_{k-1}^j\subseteq \B(\epsilon)$.
\end{proposition}

\begin{remark}
It is worth comparing the results in Theorems~\ref{theorem1}(ii) and Theorem~\ref{theorem22}.
On the one hand, the inner approximations of the minimal invariant multi-set of Theorem~\ref{theorem1} are not invariant (unless equal to the minimal invariant multi-set). whereas the outer approximations of Theorem~\ref{theorem22} always are. On the other hand, for any number $l$ of iterations of the multi-set sequence \eqref{eq_setseq_min1_1}, \eqref{eq_setseq_min1_2}, from Theorem \ref{theorem1} we always obtain an $\epsilon(l)$--approximation, which might be convenient when limited computations are allowed, whereas in Theorem~\ref{theorem22} the number of required iterations of the multi-set sequence \eqref{eq_setseq_min1_1}, \eqref{eq_setseq_min1_2} has to be larger than a threshold,  implied by \eqref{eq_prop2_1} and Theorem~\ref{theorem22}(ii). 
\end{remark}

\section{The maximal invariant multi-set} \label{section4}
First, we show that all trajectories of the System \eqref{eq_sys1}-\eqref{eq_sys3} subject to \eqref{eq_con1} converge exponentially to the minimal invariant multi-set $\{\cS_m^j\}_{j\in\cV}$.
\begin{lemma}\label{fact3}
Let $(x(\cdot),z(\cdot))$ be any solution of the System \eqref{eq_sys1}-\eqref{eq_sys3} subject to the constraints \eqref{eq_con1} and for initial conditions $z(0)\in\cV$, $x(0)\in\R^n$, $\|x \|=c>0$.
Under Assumptions~\ref{ass2}, \ref{ass3},
for any initial condition $(x(0),z(0))$ and any given $\epsilon>0$, there exists an integer $l^\star$ such that $d(x(t),\cS_{m}^{z(t)})\leq \epsilon$
for any $t\geq l^\star$, where $\{\cS_m^j\}_{j\in\cV}$ is the minimal invariant multi-set.
\end{lemma}

Given the constraint sets $\cX_j\subset\R^n$, $j\in\cV$ we define the multi-set sequence  $\{\cB_l^j\}_{j\in\cV}$, $l\geq 0$, where
\begin{align}
\cB_0^j  & =\cX_j , \quad  j\in\cV,  \label{eq_setseq_max1_1} \\
\cB_{l+1}^j& =( \bigcap\limits_{(j,d,\sigma)\in\cE}\cC(\sigma,\cB_l^d))\bigcap\cB_0^j, \ \ \ \ j\in\cV. \label{eq_setseq_max1_2}
\end{align}
The $l$-th term of the multi-set sequence \eqref{eq_setseq_max1_1}, \eqref{eq_setseq_max1_2} contains the initial conditions $(x(0),z(0))\in \cX\times \cV$ which satisfy the state constraints for at least $l$ time instants.

\begin{theorem}\label{theorem3}
Consider the System \eqref{eq_sys1}-\eqref{eq_sys3} subject to the constraints \eqref{eq_con1}, \eqref{eq_con3}. Suppose that Assumptions~\ref{ass1}-\ref{ass4} hold and let the pair $(\Gamma,\rho)$ satisfy \eqref{eq_exprel}. Moreover, assume
$\cS_m^{j}\subseteq \interior{\cX_j}$, $j\in\cV$, where $\{\cS_m^{j}\}_{j\in\cV}$ is the minimal invariant multi-set. Let
$R_j  =\max\{ R: \B(R)\subseteq \cX_j \}, \quad j\in\cV$, 
$r_j  = \min\{ r: \B(r)\supseteq \cS_m^j \}, \quad j\in\cV$,
$c = \min\{c: \B(c)\supseteq \cX_j, j\in\cV \}.$
Consider the sequence of multi-sets \eqref{eq_setseq_max1_1}, \eqref{eq_setseq_max1_2}.
The following statements hold.
\begin{itemize}
\item[(i)] There exists an integer $\s{k}\geq 1$ such that relations $\cB_{\s{k}+1}^j=\cB_{\s{k}}^j$ hold, $j\in\cV$, with
\begin{align}
\s{k}\leq  \log_{\rho}\left(\frac{\min_{j\in\cV}(R_j-r_j)}{\Gamma c}\right).\label{eq_th5_1}
\end{align}
\item[(ii)] The multi-set $\{\cB_{\s{k}}^j \}_{j\in\cV}$ is the maximal admissible invariant multi-set with respect to the System \eqref{eq_sys1}-\eqref{eq_sys3} and the constraints \eqref{eq_con1}, \eqref{eq_con3}.
\end{itemize}
\end{theorem}

\begin{remark}
We underline that the upper bound \eqref{eq_th5_1} on the number of iterations required to converge to the maximal invariant multi-set in Theorem~\ref{theorem3} can be computed a priori: The  pair $(\Gamma,\rho)$ can be recovered by applying the methods implemented in \cite{CSS:15}, the scalars $R_j, c$ depend on the problem data and can be easily computed  when the constraint sets and the disturbance sets are polyhedral or ellipsoidal sets. 
Last, the scalars $r_j$, $j\in \{1,...,M \}$ can be computed by applying the results of Section~\ref{section3}. 
\end{remark}

The relation between the maximal invariant multi-set and the maximal safe set is stated formally in the following corollary of Theorem~\ref{theorem3}. Its proof is omitted as it is straightforward.
\begin{corollary}
Consider the System \eqref{eq_sys1}--\eqref{eq_sys3} subject to the constraints \eqref{eq_con1}, \eqref{eq_con3}. Let $\{\cS_M \}_{j\in\cV}$ be the maximal invariant multi-set and let $\cY\subseteq \cV$ a set of nodes in $\cG(\cV,\cE)$. Then, the maximal safe sets $\cS_{\cY}$
with respect to the System \eqref{eq_sys1}--\eqref{eq_sys3}, the constraints \eqref{eq_con1},  \eqref{eq_con3} and with respect to $\cY\subseteq \cV$ is $\cS_\cY=\cap_{j\in\cV}\cS_M^j.$
\end{corollary}

{\color{black}{
\begin{example} \label{example33} 
We consider the example in \cite[Section 4]{MatRaph:15}, generated from modeling  possible failures of a closed-loop linear system. In \cite{MatRaph:15}, it is shown that the system is asymptotically stable, while in \cite{BenoitRaphaelParrilo:16} is was confirmed that the CJSR is precisely $\check{\rho}(\cA,\cG)= 0.9748...<1$. 
\begin{figure}[h]
\begin{center}
\begin{tikzpicture}[->,>=stealth',shorten >=1pt,auto,node distance=2.3cm, font=\small, semithick]
\tikzstyle{every state}=[fill=gray!20,draw=black,thick,text=black,scale=0.7]
\node[state,draw=black,scale=1]   (A1) {$1$};
\node[state,draw=black,scale=1]   (B1) [above left of =A1] {$2$};
\node[state,draw=black,scale=1]   (C1) [below left of =B1] {$3$};
\node[state,draw=black,scale=1]   (D1) [right of =A1] {$4$};
\path (A1) edge [bend left] node[swap] {$4$} (D1);
\path (D1) edge [bend left] node {$1$} (A1);
\path (A1) edge [loop below] node {$1$} (A1);
\path (A1) edge [bend left] node[swap] {$2$} (B1);
\path (B1) edge [bend left] node {$1$} (A1);
\path (B1) edge [bend left] node[swap] {$3$} (C1);
\path (C1) edge [bend left] node {$2$} (B1);
\path (C1) edge  node {$1$} (A1);
\path (A1) edge [bend left] node {$3$} (C1);
\end{tikzpicture}
\end{center}
\caption{The switching constraints graph $\cG(\cV,\cE)$, Example~\ref{example33}.}
\label{fig55}
\end{figure}
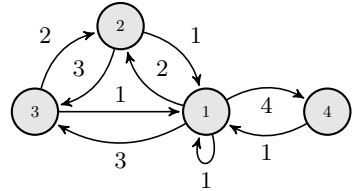
Additional to the data provided in \cite{MatRaph:15}, 
we consider state constraint sets $\cX_i=\{ x\in\R^2: \| x\|_{\infty}\leq 1\}$, $i=1,...,4$ and disturbance sets $\cW_i=0.01\cX_i$, $i=1,...,4$. We have $\cV=\{1,...,4\}$.
\begin{figure}[ht]
\psfrag{x}[][][0.8]{\color{black}{$x_1$}}
\psfrag{y}[][][0.8]{\color{black}{$x_2$}}
\psfrag{m}[][][0.8]{\color{black}{$\cS_{inn}^1$}}
\psfrag{q}[][][0.8]{\color{black}{$\cS_{inn}^2$}}
\psfrag{r}[][][0.8]{\color{black}{$\cS_{inn}^3$}}
\psfrag{v}[][][0.8]{\color{black}{$\cS_{inn}^4$}}
\psfrag{Z1}[][][0.8]{\color{black}{$\cX_1$}}
\psfrag{Z2}[][][0.8]{\color{black}{$\cX_2$}}
\psfrag{Z3}[][][0.8]{\color{black}{$\cX_3$}}
\psfrag{Z4}[][][0.8]{\color{black}{$\cX_4$}}
\psfrag{M1}[][][0.8]{\color{black}{$\cS_{M}^1$}}
\psfrag{M2}[][][0.8]{\color{black}{$\cS_{M}^2$}}
\psfrag{M3}[][][0.8]{\color{black}{$\cS_{M}^3$}}
\psfrag{M4}[][][0.8]{\color{black}{$\cS_{M}^4$}}
\begin{center}
\includegraphics[height=6cm]{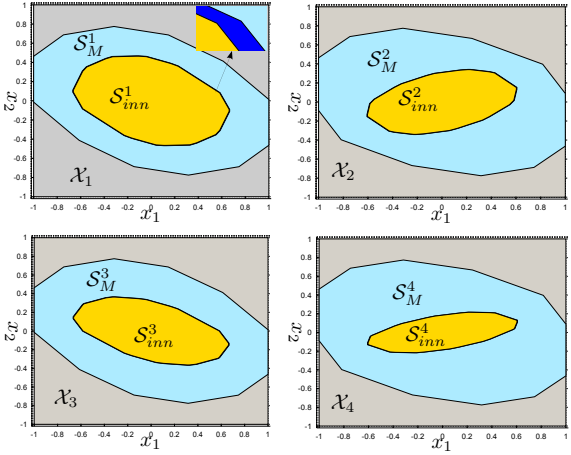}
\end{center}
\caption{ \color{black}{ Example~\ref{example33}, the convex inner $10^{-2}$--approximation  $\{\cS_{inn}^j \}_{j\in\cV}$ (yellow) of the minimal convex invariant multi-set, the maximal invariant multi-set $\{\cS_M^j \}_{j\in\cV}$ (light blue) and the constraint multi-set $\{ \cX_i\}_{i\in\cV}$ (grey).  The outer $10^{-2}$--approximation  $\{\cS_{out}^j  \}_{j\in\cV}$ is not shown clearly since its members almost coincide the ones of the inner one. In the upper left part, a zoomed portion of the inner and outer (dark blue) approximation is shown.
}}
\label{figure5}
\end{figure}
First, we compute a convex, inner $\epsilon$--approximation  $\{ \cS_{inn}^j\}_{j\in\cV}$ of the minimal convex invariant multi-set with $\epsilon=10^{-2}$ using the pair $(\Gamma,\rho)=(12.6023,0.9804)$ which satisfies \ref{eq_lemma1_1}. In Appendix B we illustrate two ways to compute such a pair. From Proposition~\ref{theorem2}, we calculate $l=328$,  thus, $\cS_{inn}^j=\s{\cF}_{328}^j$,  $j\in\cV$, where the multi-set  $\{ \s{\cF}_l^j\}_{j\in\cV}$, $l\geq 0$ is generated by the multi-set sequence  \eqref{eq_setseq_min2_1}, \eqref{eq_setseq_min2_2}. The inner $10^{-2}$-aproximation $\{\cS_{inn}^j \}_{j\in[4]}$ is shown in Figure~\ref{figure5} with yellow color.  
Second, we compute an invariant,  convex outer $10^{-2}$-approximation $\{ \cS_{out}^j\}_{j\in\cV}$ of the minimal convex invariant multi-set. By utilizing the multi-set sequences \eqref{eq_convnom1}, \eqref{eq_convnom2} and \eqref{eq_setseq_min2_1}, \eqref{eq_setseq_min2_2}, we obtain the pair $(k,\lambda)$ with $k=191$, $\lambda=0.0138$ which satisfies conditions (i) and (ii) of Proposition~\ref{proposition2c}. The convex outer $10^{-2}$-approximation is  $\{ \cS_{out}^j\}_{j\in\cV}$ with $\cS_{out}^j=\s{\cD}_{191}^j$, where $\s{\cD}_k^j$ is defined in Proposition~\ref{proposition2c}. The members of the invariant outer approximation are not clearly visible in Figure~\ref{figure5} since they almost coincide\footnote{\color{black}{Indeed, an a posteriori analysis of these two approximations shows  that they are also $\epsilon_2$--approximations with $\epsilon_2=2\cdot 10^{-3} $. In specific, we identify that $\cS_{inn}^j\subseteq\cS_{out}^j\subseteq \cS_{inn}^j\oplus \B(\epsilon_2)$. Consequently, we have that $\cS_{inn}^j\subseteq \cS_m^j\subseteq \cS_{out}^j\subseteq \cS_{inn}^j\oplus \B(\epsilon_2)\subseteq \cS_{m}^j\oplus \B(\varepsilon_2).$} }
 with the respective members of the inner convex approximation.  In the upper left part of Figure~\ref{figure5}, a zoomed portion of the image is shown and a portion of the set $\cS_{out}^1$ is shown in dark blue color. 
Last, we compute the maximal invariant multi-set $\{\cS_M^j \}_{j\in\cV}$. By utilizing Theorem~\ref{theorem3}, we compute an upper bound  $\s{k}$ on the number of iterations needed from relation \eqref{eq_th5_1}. To this purpose, 
we compute $R_j=c=1$, $j\in[4]$, $r_1=r_3=r_4=0.6671$, $r_2=0.6076$ and we have 
$\log_{\rho}(\frac{\min_{j\in\cV}(R_j-r_j)}{\Gamma c})=203$. It is worth noting that the maximal invariant multi-set is actually obtained in $\s{k}=8$ iterations. The members of the maximal invariant multi-set are shown in Figure~\ref{figure5} in light blue color.
\end{example}

}}

%
%
%

\section{The Reduced Graph and the Reduced System} \label{section5}
In this section, we show that invariant multi-sets, either maximal, minimal or their approximations, can be constructed by reachability sequences of multi-sets of graphs having a smaller number of nodes and possibly a smaller number of edges. Some of the benefits are in the safety analysis problems for systems under dwell time specifications in Section~\ref{section7}.
In specific, we  exploit the concept of the set $\cY$ of nodes of a graph $\cG(\cV,\cE)$  that are necessarily visited in any walk of length $m$, for some integer $m\geq 1$.
We note that such sets always exist with the trivial case being $\cY=\cV$, $m=1$.
\begin{definition} \emph{\textbf{( \cite{Roozbehani:08}, Proposition 1.6.7,  \cite{MatRaph:17}, Definition 4)}}
Given a graph $\cG(\cV,\cE)$ and an integer $m\geq 1$, a set of nodes $\cY\subseteq\cV$ is
called \emph{$m$-unavoidable} if any walk of length  $m$ passes through a node $v\in\cY$ at least once.
A set of nodes $\cY\subseteq\cV$
is called a \emph{minimal $m$-unavoidable set} of nodes if for any $m$-unavoidable set $\cZ\subset\R^n$, it holds that $|\cZ|\geq |\cY|$.
\end{definition}
There needs not be a unique set of $m$-unavoidable nodes, as shown in the following example.
\begin{example} \label{example4}
The graph in Figure~\ref{f1} has two minimal $2$-unavoidable sets of nodes, namely $\{a,b\}$ and $\{a,c\}$. The corresponding reduced graphs are shown in Figure~4.
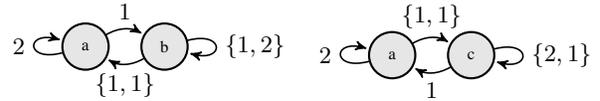
\begin{figure}[h!]
\begin{center}
\begin{tikzpicture}[->,>=stealth',shorten >=1pt,auto,node distance=1cm, font=\small, semithick]
\tikzstyle{every state}=[fill=gray!20,draw=black,thick,text=black,,scale=0.7]
\node[state,node distance=1.5cm]  (A)  {a};
\node[state,draw=black,node distance=1.5cm]         (B) [ right of =A] {b};
\path (A) edge [bend left] node {$1$} (B);
\path (A) edge [loop left] node {$2$} (A);
\path (B) edge [loop right] node {$\{1,2\}$} (B);
\path (B) edge [bend left] node {$\{1,1\}$} (A);
\end{tikzpicture}
\hspace{0.01cm}
\begin{tikzpicture}[->,>=stealth',shorten >=1pt,auto,node distance=1cm, font=\small, semithick]
\tikzstyle{every state}=[fill=gray!20,draw=black,thick,text=black,,scale=0.7]
\node[state,node distance=1.5cm]         (A)  {a};
\node[state,draw=black,node distance=1.5cm]         (B) [ right of =A] {c};
\path (A) edge [bend left] node {$\{1,1\}$} (B);
\path (A) edge [loop left] node {$2$} (A);
\path (B) edge [loop right] node {$\{2,1\}$} (B);
\path (B) edge [bend left] node {$1$} (A);
\end{tikzpicture}
\end{center}
\caption{Example \ref{example4}, the Reduced Graphs of Figure~\ref{f1} whith $\cY=\{a,b\}$ (left) and $\cY=\{a,c\}$ (right).}
\label{f5}
\end{figure}
\end{example}

Let $\cY\subseteq\cV$ be a set of $m$-unavoidable nodes of $\cG(\cV,\cE)$.
We define the graph $\cG(\cY,\cE_{\cY})$, where
\begin{align}
\cE_{\cY}  :=\{ (s,d,\sigma(s,d)): (s,d)\in\cY\times \cY \} \label{eq_reduced_graph}
\end{align}
and $\sigma(s,d)$ is a path in $\cG(\cV,\cE)$.
The edges of $\cG(\cY,\cE_{\cY})$ have as labels the sequences of labels appearing in the path in $\cG(\cV,\cE)$
from a node $s\in\cY$ to a node $d\in\cY$. 
Consider the graph $\cG(\cV,\cE)$, the set of matrices $\cA$, the set of disturbance sets $\W$, $m\in\{ 1,...,|\cV|\}$ and a set of $m$-unavoidable nodes $\cY\subseteq\cV$.
We consider the set of matrices $\tilde{\cA}\subset\R^{n\times n}$, where
$\tilde{\cA}:=\{  \prod\limits_{i=0}^p A_{\sigma_{p-i}}:  (s,d,\{\sigma_i\}_{i\in \{0,...,p \}})\in\cE_\cY  \}$. 
Let us denote each member of $\tilde{\cA}$ by $\tilde{A}_i$, $i\in\{ 1,...,\tilde{N}\}$, for some $\tilde{N}\geq 1$.
We also consider the corresponding set of disturbance sets $\tilde{\W}:=\{  \bigoplus_{j=0}^{p-1}(\prod_{i=0}^{p-1-j}A_{\sigma_{p-1-i}(t)} \cW_j) \oplus\cW_{p}: (s,d,\{\sigma_i\}_{i\in\{0,...,p \}})\in\cE_{\cY}  \}$
%
and use the notation $\tilde{\cW}_i$ for each member of $\tilde{\W}$. 
We define the \emph{Reduced Graph} $\cG(\cY,\tilde{\cE}),$
where $\cY$ is the set of unavoidable nodes and $\tilde{\cE}$ contains the same edges as $\cE_{\cY}$, \emph{with a new label} $i\in[\tilde{N}]$ for each edge corresponding to an edge $(s,d,\{\sigma_i\}_{i\in\{0,...,p \}})$ in $\cE_{\cY}$ for which the sequence $\{\sigma_i\}_{i\in\{0,...,p \}}$  has length more than one.
We emphasize the fact that  that by introducing additional modes in $\cG(\cY,\tilde{\cE})$, we have as label in each edge an integer instead of a sequence. This representation is in equivalence to the graph $\cG(\cY,\cE_\cY)$, in which the same number of modes as in $\cG(\cV,\cE)$ is kept. 
\begin{definition}[Reduced System]
Consider the System \eqref{eq_sys1}--\eqref{eq_sys3} subject to the constraints \eqref{eq_con1}--\eqref{eq_con3}. The System
\begin{align}
x(t+1) & =  \tilde{A}_{\sigma(t)}x(t)+\tilde{w}(t), \label{eq_sysred1} \\
z(t+1) & \in \Out{z(t)}{\cG(\cY,\tilde{\cE})}, \label{eq_sysred2} \\
(x(0),z(0)) & \in\R^n\times \cY, \label{eq_sysred3}
\end{align}
with $w(t)\in\tilde{\cW}_{\sigma(t)}$, subject to the constraints
\begin{align}
\sigma(t) & \in \{ \sigma: (z(t),z(t+1),\sigma)\in\tilde{\cE} \}, \label{eq_conred1} \\
x(t)&\in \cX_{z(t)}, \quad \forall t\geq 0, \label{eq_conred3}
\end{align}
is called the \emph{Reduced System}, related to the System \eqref{eq_sys1}--\eqref{eq_sys3} and the constraints \eqref{eq_con1}, \eqref{eq_con3} via the set of nodes $\cY\subseteq\cV$.
\end{definition}
The stability properties of the nominal parts of the System \eqref{eq_sys1}-\eqref{eq_sys3} and of the Reduced System \eqref{eq_sysred1}--\eqref{eq_sysred3} coincide.
\begin{Fact} \label{proposition5}
Consider the System \eqref{eq_sys1}-\eqref{eq_sys3}  and the Reduced System \eqref{eq_sysred1}--\eqref{eq_sysred3}, associated to the System via the unavoidable nodes $\cY\subseteq\cV$ and suppose that Assumption~\ref{ass3} holds. Then,  
$\check{\rho}(\tilde{\cA},\cG(\cY,\tilde{\cE}))\leq \check{\rho}(\cA,\cG(\cV,\cE)).
$
\end{Fact}

We explore the invariance relationships between the System \eqref{eq_sys1}--\eqref{eq_sys3} and the Reduced System \eqref{eq_sysred1}--\eqref{eq_sysred3}.
 Given the graph $\cG(\cV,\cE)$ and the  related Reduced Graph  $\cG(\cY,\tilde{\cE})$, we define the mapping $f(\cdot)$ from  a multi-set $\{\cM^j\}_{j\in\cY}$,
$\cM^j\subset\R^n$ to a multi-set  $\{\cK^j\}_{j\in\cV}$,   $\cK^j\subset\R^n$ to be
\begin{equation}  \label{eq_fmap}
f\left( \{\cM^j\}_{j\in\cY} \right)=\{\cK^j\}_{j\in\cV},
\end{equation}
where
\begin{equation*}
\cK^j    :=
\begin{cases}
      \cM^j, & j\in\cY, \\
      \bigcup\limits_{\{s\in\cY: m(s,j)\cap\cY=\{s\}\}} \cR(\sigma(s,j), \cM^s), &   j\in \cV\setminus \cY, 
    \end{cases}
  \end{equation*}
The set union in \eqref{eq_fmap} is over all forward reachability sets that start from a node $s\in\cY$ which is connected by a path in $\cG(\cV,\cE)$ to the node $j\in\cV\setminus \cY$ \emph{and does not pass through any other unavoidable node in $\cY$}.
We write the minimal and maximal invariant multi-set of the System and the Reduced System by
$\{ {\cS}_m^j \}_{j\in\cV}$, $\{ {\cS}_{M}^{j} \}_{j\in\cV}$
and
$\{ \tilde{\cS}_m^j \}_{j\in\cY}$, $\{ \tilde{\cS}_{M}^{j} \}_{j\in\cY}$
respectively.
Given a graph $\cG(\cV,\cE)$ and a set $\cY\subseteq\cV$, we denote by $\theta_m$ and $\theta_{M}$ the smallest and largest number of edges in a \emph{path} connecting two nodes $i\in \cY$, i.e.,
\begin{align}
\theta_m & := \min\limits_{(i,j)\in\cY\times \cY}\{ \gd{i}{j} \}, \label{eq_thsmall}\\
\theta_M & := \max\limits_{(i,j)\in\cY\times \cY}\{ \gd{i}{j} \}. \label{eq_thlarge}
\end{align}
The following Lemma states that the $l$-step forward reachability multi-set of the Reduced System is bounded from above and below from by the $l\cdot \theta_M$-step and $l\cdot\theta_m$-step forward reachability multi-sets of the original System. 
\begin{lemma}\label{lemma4}
Let $\{\cF_{l}^j \}_{j\in\cV}$, $\{\tilde{\cF}_l^j\}_{l\in\cY}$ be the forward reachability multi-set sequences \eqref{eq_setseq_min1_1}, \eqref{eq_setseq_min1_2}
of the System \eqref{eq_sys1}--\eqref{eq_sys3} and the Reduced System \eqref{eq_sysred1}--\eqref{eq_sysred3} associated to the System via the set of nodes $\cY$ respectively.
Then,
\begin{align}\label{eq_lem4_1}
\cF_{l\theta_m}^j\subseteq \tilde{\cF}_{l}^j\subseteq \cF_{l\theta_M}^j,\quad \forall j\in\cY,\quad \forall l\geq 0,
\end{align}
where $\theta_m$, $\theta_M$ are given in \eqref{eq_thsmall} and \eqref{eq_thlarge} respectively.
\end{lemma}

\begin{theorem}\label{theorem6}
Consider the System \eqref{eq_sys1}--\eqref{eq_sys3} subject to the constraints \eqref{eq_con1}--\eqref{eq_con3}, a set of unavoidable nodes $\cY\subseteq \cV$, the Reduced System
\eqref{eq_sysred1}--\eqref{eq_sysred3} subject to the constraints \eqref{eq_conred1}, \eqref{eq_conred3} related to the System via the set of nodes $\cY$, with the minimal invariant multi-set $\{ \tilde{\cS}_m^j \}_{j\in\cY}$. 
 Under Assumptions \ref{ass1}--\ref{ass4}, the minimal invariant multi-set $\{ \cS_m^j\}_{j\in\cV}$ with respect to the System \eqref{eq_sys1}--\eqref{eq_sys3} and the constraints \eqref{eq_con1}--\eqref{eq_con3}  is
\begin{equation}\label{eq_th4_1}
 \{\cS_m^j\}_{j\in\cV}=f\left(\{ \tilde{\cS}_m^j \}_{j\in\cY}\right).
\end{equation}
\end{theorem}
Theorem~\ref{theorem6} reveals the relationship between the minimal invariant multi-sets of the Reduced System and the original System. In what follows, we show that the relation extends also to their $\epsilon$--approximations, which is especially appealing in formulating efficient algorithmic procedures for their computation.

\begin{Fact} \label{fact2}
Consider a system of the form \eqref{eq_sys1}--\eqref{eq_sys3} subject to the constraints \eqref{eq_con1}, \eqref{eq_con3}
the sequence $\{\sigma_i\}_{i\in\{1,...,l \}}$, $\sigma_i\in\{1,...,N \}$ and two sets $\cS_1,\cS_2\subset\R^n$.
Then, for all $l\geq 1$ it holds that
\begin{equation}\label{eq_fact2_1}
\cR(\{\sigma_i  \}_{i\in[l]},\cS_1\oplus\cS_2)=\cR(\{\sigma_i  \}_{i\in[l]},\cS_1)\oplus\cR_{\text{N}}(\{\sigma_i  \}_{i\in[l]},\cS_2).
\end{equation}
\end{Fact}
The sets $\cS_1$, $\cS_2$ in \eqref{eq_fact2_1} of Fact~\ref{fact2} commute.
Similarly to the multi-set $\{ \cN_{l}^j \}_{j\in\cV}$ generated by \eqref{eq_setseq1_1}, \eqref{eq_setseq1_2}, we define the sequence $\{ \tilde{\cN}_{l}^j  \}_{j\in\cY}$, $l\geq 0$, with $\tilde{\cN_0}^j=\bigcup\limits_{
(i,j,\sigma)\in\tilde{\cE}}\tilde{\cW}_i$, $\quad \tilde{\cN}_{l+1}^j=\bigcup\limits_{(i,j,\sigma)\in\tilde{\cE}} \cR_{\text{N}}(\sigma,\tilde{\cN}_l^i)$.
Moreover, we define
\begin{equation}\label{eq_need_1}
\tilde{\alpha}=\min\{ \alpha: \cup_{\sigma\in \{1,...,\tilde{N}\}}\tilde{\cW}_{\sigma}\subseteq \B(\alpha) \}.
\end{equation}
By Assumption~\ref{ass3}, there exist scalars $\tilde{\Gamma}\geq 1$, $\tilde{\rho}<1$
such that
\begin{equation}\label{eq_need_2}
\tilde{\cN}_t^j\subseteq \tilde{\Gamma}\tilde{\rho}^t \tilde{\cN}_0.
\end{equation}
Additionally, since the nominal part of the System \eqref{eq_sys1}--\eqref{eq_sys3}
is exponentially stable, there exist scalars $\Gamma\geq 1$, $\rho\in (0,1)$
such that \eqref{eq_exprel} holds. Expressing \eqref{eq_exprel} using the forward reachability multi-sets of the nominal system, for any admissible switching sequence $\sigma(i,j)$, $(i,j)\in\cV\times \cV$ we write
\begin{equation}\label{eq_need3}
\cR_{\text{N}}(\sigma(i,j),\B(1))\subseteq \Gamma \rho^{|\sigma(i,j)|}\B(1).
\end{equation}

\begin{proposition}\label{theorem7}
Consider the System \eqref{eq_sys1}--\eqref{eq_sys3} subject to the constraints \eqref{eq_con1},\eqref{eq_con3}, a set of unavoidable nodes $\cY\subseteq \cV$, the associated Reduced System
\eqref{eq_sysred1}--\eqref{eq_sysred3} subject to the constraints \eqref{eq_conred1}, \eqref{eq_conred3} and the multi-set sequence $\{ \tilde{\cF}_l^j \}_{j\in\cV}$, $l\geq \theta_M$, with
$\{\tilde{\cF}_{l}^j\}_{j\in\cV}=f(\{ \tilde{\cF}_{l}^j \}_{j\in\cY})$, where $f(\cdot)$ is given by \eqref{eq_fmap} and $\{\tilde{\cF}_{l}^j\}_{j\in\cY}$ is generated by the multi-set sequence $\tilde{\cF}_0^j=\{0\}$, $\tilde{\cF}_{l+1}^j=\bigcup\limits_{(i,j,\sigma)\in\tilde{\cE}}\cR(\sigma,\tilde{\cF}_l^i)$, for all $l\geq 0$, for all $j\in\cY$.
Then, for any $\epsilon>0$ and
\begin{equation} \label{eq_th7_1}
l\geq \max\left\{ \theta_M,\left\lceil \log_{\tilde{\rho}} \left( \frac{\epsilon(1-\tilde{\rho})}{\tilde{\alpha} \tilde{\Gamma}\max\{1,\Gamma \rho \} }\right)  \right\rceil \right\},
\end{equation}
where $\tilde{\alpha}, \tilde{\Gamma}, \tilde{\rho}, \Gamma, \rho$ are defined by \eqref{eq_need_1}--\eqref{eq_need3}, it holds that
\begin{equation}\label{eq_th7_2}
\tilde{\cF}_l^j\subseteq \cS_m^j\subseteq \tilde{\cF}_l^j\oplus\B(\epsilon), \quad \forall j\in\cV.
\end{equation}
\end{proposition}

\begin{proposition}\label{theorem8}
Consider the System \eqref{eq_sys1}--\eqref{eq_sys3} subject to the constraints \eqref{eq_con1}, \eqref{eq_con3}, a set of unavoidable nodes $\cY\subseteq \cV$, the associated Reduced System
\eqref{eq_sysred1}--\eqref{eq_sysred3} subject to the constraints \eqref{eq_conred1}, \eqref{eq_conred3}, the multi-set sequence $\tilde{\cF}_0^j=\{0\}$, $\tilde{\cF}_{l+1}^j=\bigcup\limits_{(i,j,\sigma)\in\tilde{\cE}}\cR(\sigma,\tilde{\cF}_l^i)$, for all $l\geq 0$, for all $j\in\cY$, and a positive scalar $\epsilon>0$.
Consider the multi-set $\{\tilde{\cD}_k^j \}_{j\in\cV}=f(\frac{1}{1-\lambda}\{\tilde{\cF}_{k-1}^j \}_{j\in\cY})$, where $f(\cdot)$ is defined in \eqref{eq_fmap} and the pair $k\geq 1$, $\lambda\in(0,1)$ satisfies
\begin{align}
\tilde{\cN}_k^j & \subseteq \lambda \tilde{\cN}_\cap^{j}, \quad \forall  j\in\cY, \label{eq_th8_1} \\
\frac{\lambda \max\{\Gamma \rho,1\}}{1-\lambda}\tilde{\cF}_{k-1}^j & \subseteq \B(\epsilon), \quad \forall j\in\cY, \label{eq_th8_2} \\
k & \geq \theta_M \label{eq_th8_3},
\end{align}
where $\Gamma, \rho$ satisfy \eqref{eq_need3} and $\tilde{\cN}_{\cap}^j=\cup_{(i,j,\sigma\in\cY)}\tilde{\cW}_{\sigma}$, $j\in\cY$.
Then, the multi-set $\{\tilde{\cD}_{k}^{j}  \}_{j\in\cV}$ is an invariant, outer $\epsilon$-approximation of the minimal invariant multi-set, i.e.,
\begin{equation}\label{eq_th8_4}
\cS_m^j\subseteq \tilde{\cD}_{k}^j\subseteq \cS_m^j\oplus \B(\epsilon), \quad \forall j\in\cV.
\end{equation}
\end{proposition}

Propositions \ref{theorem7} and \ref{theorem8} enable us to compute the minimal invariant multi-set and its $\epsilon$--approximations by performing the  computations in the Reduced System and the Reduced Graph. This is not in general the case when computing the maximal admissible invariant multi-set. More precisely, the requirement that state constraints need to be satisfied at the non-unavoidable nodes does not allow the forward propagation of the elements of the multi-set from the unavoidable nodes, as it is the case in Theorem~\ref{theorem6}. However, an explicit relation between the maximal admissible multi-sets of the Reduced System and the System does exist for some interesting  cases, as it is shown in Section \ref{section7}.  Generally, the Reduced System offers an outer approximation of the maximal invariant multi-set of the System. This is stated formally below, without proof.

\begin{proposition}\label{theorem9}
Consider the System \eqref{eq_sys1}--\eqref{eq_sys3} subject to the constraints \eqref{eq_con1}--\eqref{eq_con3}, a set of unavoidable nodes $\cY\subseteq \cV$, the Reduced System
\eqref{eq_sysred1}--\eqref{eq_sysred3} subject to the constraints \eqref{eq_conred1}, \eqref{eq_conred3}.
Let $\{\tilde{\cS}_M^j\}_{j\in\cY}$ and $\{\cS_M^j\}_{j\in\cV}$ denote the maximal admissible invariant multi-set of the System and the Reduced System respectively.
Then, under Assumptions \ref{ass1}--\ref{ass4}, the following hold: (i) $\cS_M^j\subseteq\tilde{\cS}_M^j$, for all $j\in\cY$. (ii) Consider the sets $\cB_0^j=\tilde{\cS}_M^j$, $j\in\cY$, $\cB_0^j=\cX_j$ and $j\in\cV\setminus \cY$. Then the multi-set sequence generated by \eqref{eq_setseq_max1_1}, \eqref{eq_setseq_max1_2} converges in finite time to the maximal invariant multi-set of the System.
\end{proposition}

\section{The Lifted Graph and the Lifted System} \label{section6}
In this section, we apply two types of lifting procedures on the graph $\cG(\cV,\cE)$
that defines the switching constraints on the System \eqref{eq_sys1}--\eqref{eq_sys3}.
More specifically, we expose the relation between the minimal and maximal invariant multi-sets between the System and the Lifted System, which can be exploited in several different ways.

\subsection{The T-product Lift}
We consider the T-iterated dynamics of the System.
This relaxation has been introduced before for assessing stability, e.g., in \cite{Peut:98}
for continuous-time systems, in \cite{Mircea13} for homogeneous discrete-time systems and
\cite{Roman:14} for non-linear difference equations. In the context of linear constrained switching systems, the T-product Lift has been introduced in \cite{MatRaph:15}
where it was shown it provides asymptotically tight approximations of the CJSR.

\begin{definition}[T-Lift \cite{MatRaph:15}]
Consider the System \eqref{eq_sys1}--\eqref{eq_sys3} subject to the constraints \eqref{eq_con1}--\eqref{eq_con3} and the related switching constraints graph $\cG(\cV,\cE)$.
Given an integer $T\geq 1$, the T-product lifted graph $\cG_T(\cV,\cE_T)$ is a graph having the same nodes with $\cG(\cV,\cE)$
and  the set of edges
$\cE_T:=\{ (i,j,\sigma(i,j)): (i,j)\in\cV\times \cV, |\sigma(i,j)|=T \}.$
\end{definition}
The T-lifted graph $\cG_T(\cV,\cE_T)$ has the same number of nodes with $\cG(\cV,\cE)$. Moreover, there is an edge between a node $i\in\cV$ and $j\in\cV$ whenever there is a 
walk between $i$ and $j$ in $\cG(\cV,\cE)$ of length $T$.
Let $\cA^T:=\{ \prod\limits_{i=1}^T A_{\sigma_{T-i}}:\left( \exists
\sigma_i\in \{1,...,N \}, {i\in \{1,...,T \}} :  (i,j,\sigma(i,j))\in\cE_T \right.$ and $ \left.  \{\sigma_i\}_{i\in\{1,...,T \}}=\sigma(i,j) \right)	 \}$
be the set of matrices formed by the products corresponding to labels appearing in a walk of length $T$ in the graph $(\cG,\cV)$ and
$\mathbb{W}^T:=\{  \bigoplus_{i=1}^{T}(\prod_{j=1}^{T-i}A_{\sigma_{T-j}} \cW_i):  ( \exists
\sigma_l\in\{1,...,N \}, l\in\{1,...,T \}:  (s,d,\sigma(s,d))\in\cE_T, \text{ and } \{\sigma_l\}_{l\in\{1,...,T \}}=\sigma(s,d) )	  \}$
be the corresponding set of disturbance sets of the iterated dynamics.
\begin{example} \label{example5}
The $2$-product Lifted Graph of the Graph of Example~\ref{example1} is shown in Figure~\ref{f6}. The new labels $\check{\sigma_1},...,\check{\sigma}_4$ correspond to the switching sequences $\{1,1\}, \{1,2 \}, \{2,1 \}$ and $\{2,2 \}$ respectively.
 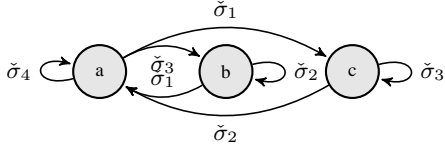
\begin{figure}[ht]
\begin{center}
\begin{tikzpicture}[->,>=stealth',shorten >=1pt,auto,node distance=2.1cm, font=\small, semithick]
\tikzstyle{every state}=[fill=gray!20,draw=black,thick,text=black,,scale=0.8]
\node[state,node distance=2.2cm]         (A)  {a};
\node[state,draw=black]         (B) [ right of =A] {b};
\node[state,draw=black]         (C) [ right of =B] {c};
\path (A) edge [loop left] node {$\check{\sigma}_{4}$} (A);
\path (B) edge [loop right] node {$\check{\sigma}_2$} (B);
\path (C) edge [loop right] node {$\check{\sigma}_3$} (C);
\path (A) edge [bend left] node [swap] {$\check{\sigma}_3$} (B);
\path (A) edge [bend left] node {$\check{\sigma}_1$} (C);
\path (B) edge [bend left] node [ swap] {$\check{\sigma}_1$} (A);
\path (C) edge [bend left] node {$\check{\sigma}_2$} (A);
\end{tikzpicture}
\end{center}
\caption{Example \ref{example5}, the $2$-product Lifted Graph of the graph of Figure~\ref{f1}.
}
\label{f6}
\end{figure}
\end{example}
\begin{definition}[T-product Lifted System]
Consider the System \eqref{eq_sys1}--\eqref{eq_sys3} subject to the constraints \eqref{eq_con1}--\eqref{eq_con3} and an integer $T\geq 1$.
The System
\begin{align}
x(t+1)= A_{T,\sigma(t)}x(t)+w_T(t)  \label{eq_liftsys1_1}\\
z(t+1)\in \Out{z(t)}{\cG_T}, \label{eq_liftsys1_2}		\\
(x(0),z(0))\in\R^n\times \cV, \label{eq_liftsys1_3}
\end{align}
with $w(t)\in \cW^T_{\sigma(t)}\in\mathbb{W}^T$, $A_{T,\sigma(t)}\in\cA^T$, subject to the constraints
\begin{align}
\sigma(t) \in\{\sigma: (z(t),z(t+1),\sigma)\in\cE_T \}, \label{eq_liftcon1_1} \\
x(t)\in\cX_{z(t)}, \quad \forall t\geq 0, \label{eq_liftcon1_2}
\end{align}
is called the T-product Lifted System, related to the System \eqref{eq_sys1}--\eqref{eq_sys3}
and the constraints \eqref{eq_con1}--\eqref{eq_con3}.
\end{definition}
It has been shown that the asymptotic stability properties of the System \eqref{eq_sys1}--\eqref{eq_sys3} and the Lifted System \eqref{eq_liftsys1_1}--\eqref{eq_liftsys1_3}  coincide \cite[Theorem 3.2]{MatRaph:15}.
and that the CJSR of the T-product Lifted System is equal to
the T-th power of the CJSR of the System.
Here, we reveal the relationship between the minimal and maximal invariant multi-sets of the System and the Lifted System and we use the latter fact to propose alternative algorithmic procedures to compute them.

\begin{theorem}\label{theorem99}
Consider the System \eqref{eq_sys1}--\eqref{eq_sys3} subject to the constraints \eqref{eq_con1}--\eqref{eq_con3}, an integer $T\geq 0$
and the  $T$-product Lifted System \eqref{eq_liftsys1_1}--\eqref{eq_liftsys1_3} subject to the constraints \eqref{eq_liftcon1_1}, \eqref{eq_liftcon1_2}.
Let $\{ \check{\cS}_m^j\}_{j\in\cV}$, $\{ \check{\cS}_M^j\}_{j\in\cV}$ be the minimal and the maximal invariant multi-set with respect to the T-product Lifted System. The following statements hold:
\noindent (i) The multi-set 
\begin{align}    \resizebox{0.88\hsize}{!}{%
$
\cS^j = \check{\cS}_m^j\bigcup\limits_{ \{  s\in\cV: |\sigma(s,j)|\leq T-1   \}}\cR(\sigma(s,j), \check{\cS}_m^s), \quad j\in\cV, \label{eq_th9_1}$
}
\end{align}
is the minimal invariant multi-set with respect to the System, i.e., $\cS^j=\cS_m^j$, for all $j\in\cV$.
\noindent (ii) The multi-set
\begin{align}
\resizebox{0.88\hsize}{!}{%
$
\cS^j = \check{\cS}_M^j\bigcap\limits_{ \{  d\in\cV: |\sigma(s,j)|\leq T-1   \}}\cC(\sigma(j,d), \check{\cS}_M^d), \quad j\in\cV, \label{eq_th10_1}
$
}
\end{align}
is the maximal invariant multi-set  with respect to the System, i.e., $\cS^j=\cS_M^j$, for all $j\in\cV$.
\end{theorem}

\subsection{The P-Path-Dependent Lift}
We explore the potential benefits of computing invariant multi-sets for another type of lifting. In specific, we study the Path-Dependent lifting, introduced in \cite{BliTre:03} and \cite{LeeDu:06}. Relevant to the setting studied here, the lifting has been used for the stability
 analysis of constrained switching systems in \cite{MatRaph:15} where it was shown to provide asymptotically tight approximations to the constrained joint spectral radius.

\begin{definition}[P-Lift \cite{LeeDu:06}]
Consider an integer $P\geq 1$ and a graph $\cG(\cV,\cE)$, that corresponds to a System \eqref{eq_sys1}--\eqref{eq_sys3} subject to the constraints \eqref{eq_con1}--\eqref{eq_con3}.
The Path dependent lifted graph $\cG_P(\cV_P,\cE_P)$ is a graph  with the set of nodes $\cV_P$, 
$\cV_P:=\{ v_{i_1}\sigma_{i_1}v_{i_2}\cdots v_{i_{P+1}}: (v_{i_j},v_{i_{j+1}},\sigma_{i_j})\in\cE, j\in\{1,...,P \} \},
$
and the set of edges 
$\cE_P:=\{  (v_a,v_b,\sigma): (\exists v_{i_j}\in\cV, j\in\{1,...,P+2 \}:  v_a=v_{i_1}\sigma_{i_1}\cdots v_{i_{P+1}}, v_b=v_{i_2}\sigma_{i_2}\cdots v_{i_{P+2}}, \sigma=\sigma_{i_{P+1}}, \sigma_{i_j}\in\{1,...,N \}  ) \}.$
\end{definition}
Moreover, for each node $v_a=v_{i_1}\sigma_{i_1}\cdots v_{i_{P+1}} \in\cE_P$ we assign the constraint set $\cX_a=\cX_{i_{P+1}}$.
Roughly, the P-Path-Dependent Lifted graph $\cG_P(\cV_P,\cE_P)$ has as many nodes as different walks of length $P-1$ in the graph $\cG(\cV,\cE)$. Moreover, there is an edge from  node $s=v_{i_1}\sigma_{i_1}\cdots v_{i_{P+1}}\in\cV_P$ and to a node $d=v_{l_1}\sigma_{l_1}\cdots v_{l_{P+1}}\in\cV_P$  if $v_{i_2}\sigma_{i_2}\cdots v_{i_{P+1}}=v_{l_1}\sigma_{l_1}\cdots v_{l_P}$.

\begin{example} \label{example6}
The $1$-Path-Dependent Lifted Graph of the Graph in Example~\ref{example1} is shown in Figure~\ref{f7}.
\begin{figure}[ht]
\begin{center}
\begin{tikzpicture}[->,>=stealth',shorten >=1pt,auto,node distance=1.5cm, font=\small, semithick]
\tikzstyle{every state}=[fill=gray!20,draw=black,thick,text=black,,scale=0.8]
\node[state,node distance=1.5cm]         (A)  {a2a};
\node[state,draw=black]         (B) [ right of =A] {a1b};
\node[state,draw=black]         (C) [ right of =B] {b1c};
\node[state,draw=black]         (D) [ right of =C] {c2b};
\node[state,draw=black]         (E) [ right of =D] {c1a};
\path (A) edge [loop left] node {$2$} (A);
\path (A) edge [bend left] node  {$1$} (B);
\path (B) edge [bend left] node  {$1$} (C);
\path (C) edge [bend left] node [swap] {$2$} (D);
\path (C) edge [bend left] node  {$1$} (E);
\path (D) edge [bend left] node  {$1$} (C);
\path (E) edge [bend left] node  {$2$} (A);

\end{tikzpicture}
\end{center}
\caption{Example \ref{example6}, the $2$-product Lifted Graph of the graph of Figure~\ref{f1}.
}
\label{f7}
\end{figure}
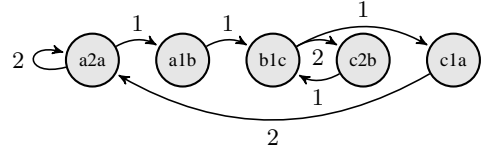
\end{example}

\begin{definition}[P-Path-Dependent Lifted System]
Consider the System \eqref{eq_sys1}--\eqref{eq_sys3} subject to the constraints \eqref{eq_con1}, \eqref{eq_con3} and an integer $P\geq 1$. The System
\begin{align}
x(t+1)= A_{\sigma(t)}x(t)+w(t)  \label{eq_liftsys2_1}\\
z(t+1)\in \Out{z(t)}{\cG_P}, \label{eq_liftsys2_2}		\\
(x(0),z(0))\in\R^n\times \cV_P, \label{eq_liftsys2_3}
\end{align}

with $w(t)\in \cW_{\sigma(t)}$, $A_{\sigma(t)}\in\cA$, subject to the constraints
\begin{align}
\sigma(t) \in\{\sigma: (z(t),z(t+1),\sigma)\in\cE_P \}, \label{eq_liftcon2_1} \\
x(t)\in\cX_{z(t)}, \quad \forall t\geq 0, \label{eq_liftcon2_2}
\end{align}
is called the P-Path-Dependent Lifted System, related to the System \eqref{eq_sys1}--\eqref{eq_sys3}
and the constraints \eqref{eq_con1}--\eqref{eq_con3}.
\end{definition}
For any $j\in\cV$, we define the  sets of nodes  $\cI(j)\subseteq \cV_P$, $j\in |\cV|$, where
$\cI(j):=\{ i\in\cV_P:  i=v_{i_1}\sigma_{i_1}\cdots j: (v_{i_j}, v_{i_{j+1}},\sigma_{i_j})\in \cE, j\in\{1,...,P-1 \}, (v_{i_P},j,\sigma_{i_P})\in\cE   \}.  \label{eq_th11_1}
$

\begin{theorem}\label{theorem11}
Consider the System \eqref{eq_sys1}--\eqref{eq_sys3} subject to the constraints \eqref{eq_con1}--\eqref{eq_con3}, an integer $P\geq 0$
and the  $P$-Path-Dependent Lifted System \eqref{eq_liftsys2_1}--\eqref{eq_liftsys2_3} subject to the constraints \eqref{eq_liftcon2_1}, \eqref{eq_liftcon2_2}.
Let $\{ \check{\cS}_m^j\}_{j\in\cV_P}$ and  $\{ \check{\cS}_M^j\}_{j\in\cV_P}$ be the minimal and maximal invariant multi-set with respect to the P-Path-Dependent Lifted System respectively. The following statements hold. (i) The multi-set
\begin{align}
\cS^j =   \bigcup\limits_{ i\in\cI(j)}  \check{\cS}_m^i, \quad  \quad j\in\cV, \label{eq_th11_2}
\end{align}
where $\cI(j)$ is defined in \eqref{eq_th11_1}, is the minimal invariant multi-set with respect to the System, i.e., $\cS^j=\cS_m^j$, for all  $j\in\cV$.
(ii) The multi-set
\begin{align}
\cS^j =  \check{\cS}_M^i, \quad i\in\cI(j),  \label{eq_th12_1}
\end{align}
where $\cI(j)$ is defined in \eqref{eq_th11_1}, is the maximal invariant multi-set with respect to the System, i.e., $\cS^j=\cS_M^j$, for all $j\in\cV$.
\end{theorem}

\begin{remark}\label{remark_cortheorem6}
Building on the proof of Theorem~\ref{theorem11}(i), we can show that the relation \eqref{eq_th11_2} can be extended to analogous statements concerning the minimal convex invariant multi-set and the $\epsilon$--approximations of the minimal invariant multi-set.
\end{remark}

\section{Applications}\label{section7}

\subsection{Switching under dwell time restrictions} \label{section_app1}

We apply the results of Section~\ref{section5} to two interesting control applications that concern dwell time specifications. In specific, we compute $\epsilon$-approximations of the minimal invariant multi-set via the Reduced Graph and we further refine Proposition~\ref{theorem9} to compute exactly the maximal invariant multi-set from the Reduced system, for the special cases of minimum/maximum dwell time specifications.

\subsubsection{Minimum dwell time}
We consider switching constraints which
impose a restriction on how fast switching from one mode to another is possible. In specific, given a set of $N$ modes, $N>1$, and a dwell time $\tau>1$,
the dynamics of the system may switch from a mode $i\in\{1,...,N \}$ to another mode $j\in\{1,...,N \}$ only if the system has followed the dynamics of the mode $i$ for at least $\tau$ consecutive time instants.
This type of switching constraints can be described by a graph $\cG(\cV,\cE)$ with
$|\cV|=N(N-1)(\tau-1)+N$ nodes and $|\cE|=N(N-1)\tau+N$ edges.
 For example, when $N=2$,
 a graph $\cG(\cV,\cE)$ that captures the minimum dwell-time constraints is shown in Figure~\ref{fig55}. Thus, expressing the switching system as a System \eqref{eq_sys1}--\eqref{eq_sys3} subject to constraints \eqref{eq_con1}, \eqref{eq_con3} is
 possible.
\begin{remark}
To deal with the dwell-time constraints at $t=0$, one can either pose  the \emph{additional constraint} that $z(0)$ is in the set of the unavoidable nodes of the graph $\cG(\cV,\cE)$.
\end{remark}
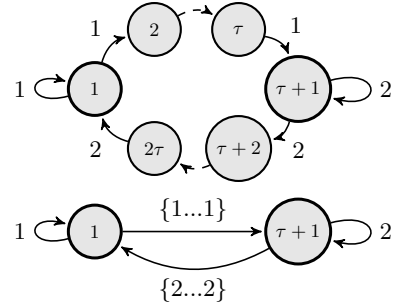
\begin{figure}[h]
\begin{center}
\begin{tikzpicture}[->,>=stealth',shorten >=1pt,auto,node distance=1.4cm, font=\small, semithick]
\tikzstyle{every state}=[fill=gray!20,draw=black,thick,text=black,scale=1]
\node[state,draw=black,scale=0.8, line width=0.4mm]         (A)  {$1$};
\node[state,draw=black,scale=0.8]         (B) [ above right of =A] {$2$};
\node[state,draw=black,scale=0.8]         (C) [ right of =B] {$\tau$};
\node[state,draw=black,scale=0.8, , line width=0.4mm]         (D) [below right of =C] {$\tau+1$};
\node[state,draw=black,scale=0.8]         (E) [below left of =D] {$\tau+2$};
\node[state,draw=black,scale=0.8]         (G) [ left of =E] {$2\tau$};
\path (A) edge [bend left] node {$1$} (B);
\path (B) edge [bend left,dashed] node {} (C);
\path (C) edge [bend left] node {$1$} (D);
\path (D) edge [bend left] node {$2$} (E);
\path (E) edge [bend left,dashed] node {} (G);
\path (G) edge [bend left] node {$2$} (A);
\path (D) edge [loop right] node {$2$} (D);
\path (A) edge [loop left] node {$1$} (A);
\end{tikzpicture}
\hspace{0.5cm}
\begin{tikzpicture}[->,>=stealth',shorten >=1pt,auto,node distance=1.4cm, font=\small, semithick]
\tikzstyle{every state}=[fill=gray!20,draw=black,thick,text=black,scale=1]
\node[state,draw=black,scale=0.8, line width=0.4mm]         (A)  {$1$};
\node[state,draw=black,scale=0.8, , line width=0.4mm]         (D) [below right of =C] {$\tau+1$};
\path (A) edge  node {$\{1...1\}$} (D);
\path (D) edge [bend left] node {$\{2...2\}$} (A);
\path (D) edge [loop right] node {$2$} (D);
\path (A) edge [loop left] node {$1$} (A);
\end{tikzpicture}
\end{center}
\caption{A minimum dwell-time constraints graph $\cG(\cV,\cE)$ (above), for a system consisting  of two nodes, with a dwell-time $\tau>1$ and the Reduced Graph taking $\cY=\{1,\tau+1 \}$ (below).}
\label{fig55}
\end{figure}
The smallest set of unavoidable nodes is unique and consists of $N$ nodes (e.g., in Figure~\ref{fig55} we have $\cY=\{1,\tau+1 \}$).

\begin{proposition}\label{theorem13}
Consider a system with $N$ modes, subject to minimum dwell-time constraints with $\tau\geq 1$, expressed in the form of the System \eqref{eq_sys1}--\eqref{eq_con3} and with a common state constraint set $\cX_j=\cX\subset\R^n$, $j\in\cV$.
Moreover, consider the set of unavoidable nodes $\cY\subseteq \cV$, the Reduced System
\eqref{eq_sysred1}--\eqref{eq_sysred3} subject to the constraints \eqref{eq_conred1}, \eqref{eq_conred3}.
Let $\{ \tilde{\cS}_M^j \}_{j\in\cY}$ be the maximal invariant multi-set with respect to the Reduced System.
 Under Assumptions \ref{ass1}--\ref{ass3}, the maximal invariant multi-set $\{\cS_M^j \}_{j\in\cV}$ with respect to the System \eqref{eq_sys1}--\eqref{eq_sys3} and the constraints \eqref{eq_con1}--\eqref{eq_con3}  is given by \eqref{eq_th13_1}.
\begin{figure*}[!t]
 \begin{equation} \label{eq_th13_1}
\cS^j_M    :=
\begin{cases}
      \tilde{\cS}_M^j, & j\in\cY, \\
\bigcap\limits_{\{d\in\cY: m(j,d)\cap\cY=\{d \} \}} \left(    \left(	\bigcap\limits_{\{i\in m(j,d)\setminus \{j,d\} \}}	\cC(\sigma(j,i), \cX  )	\right) \cap \cC(\sigma(j,d),\tilde{\cS}_M^d) 	\right)\cap \cX, &   j\in \cV\setminus \cY.
    \end{cases}
  \end{equation}
  \end{figure*}

\end{proposition}

For the studied case, the reduced graph $\cG(\cY,\hat{\cE})$ is a fully connected graph consisting of $N$ nodes and $N^2$ edges, which are significantly less than the $N(N-1)(\tau-1)+N$ nodes and $N(N-1)\tau+N$ edges of the original graph $\cG(\cV,\cE)$.

\begin{example}\label{example7}
\begin{table*}[t]
\label{table2}
\color{black}{
\centering
\begin{tabular}{|c|c|c|c|c|c|c|c|c|c|c|c|} \hline
$\tau$ & $t_{inn}(s)$ &  $\#_{inn}$ & $l_{in}$ & $t_{out}(s)$ &  $\#_{out}$ & $l_{out}$ & $t_{max}(s)$ &  $\#_{max}$ & $\s{k}$   \\ \hline

$6$ & $7.17$ & $42$ & $246$ & $1.92$ & $42$ & $147$ & $0.07$ & $14$ & $8$ \\ \hline

$10$ & $6.56$ & $56$ & $119$ & $1.26$ & $56$ & $60$ & $0.07$ & $14$ & $7$ \\ \hline

\end{tabular}
\caption{\color{black}{Example~\ref{example7}. The integer $\tau$ is the minimum dwell time, $t$ is the time required for the computations in seconds, $l$ is the upper bound on the iterations required for reaching the desired accuracy and $\#_{\_}$ stands for the maximum number of vertices that a member of the respective multi-set has. The subscripts $inn$, $out$ and $max$ stand for the inner approximation of the convex minimal invariant multi-set, the outer approximation  of the convex minimal invariant multi-set and the maximal invariant multi-set. The integer $\s{k}$ is the number of iterations required to reach the maximal invariant multi-set. }
}}
\end{table*}
{\color{black}{We consider the two-dimensional systems considered in \cite[Section 6, Systems Ia, Ib]{DehgOng:12b}. Therein, the concepts of the minimal and maximal Disturbance Dwell-Time (DDT) invariant sets were introduced for systems under minimum dwell time restrictions. The main idea was to transform the constrained switching system in an arbitrary switching consisting of $N\cdot \tau$ modes, where $\tau$ is the minimum dwell time and $N$ the number of  the initial modes. In our setting, the maximal DDT set is equal to the maximal safe set with respect to the unavoidable set of nodes, as defined in Corollary~1. 
 Utilizing the Reduced System via Propositions~\ref{theorem7} and \ref{theorem8}, we compute the $\epsilon$--inner and $\epsilon$--outer approximations of the convex minimal invariant multi-set, for $\epsilon=10^{-2}$. Consequently, we can compute the $\epsilon$--approximations of the minimal DDT set\footnote{It coincides with the intersection of the members of the minimal invariant multi-set which correspond to the unavoidable nodes.} in \cite{DehgOng:12b}, \emph{which was not possible before}. Moreover, it is worth observing that the computational times are much shorter compared to \cite{DehgOng:12b}. 
The maximal invariant multi-set can also be computed utilizing the Reduced System via Proposition~\ref{theorem13}. In Table 3, all the respective computation times, the number of iterations and the complexity of the representation of the multi-sets are shown.
}}
\end{example}

\subsubsection{Maximum dwell time}
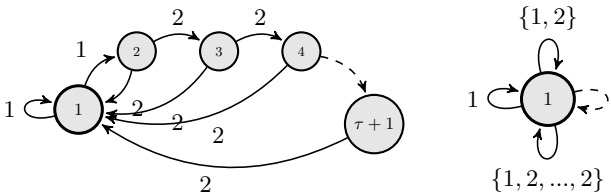
\begin{figure}[h]
\begin{center}
\begin{tikzpicture}[->,>=stealth',shorten >=1pt,auto,node distance=1.9cm, font=\small, semithick]
\tikzstyle{every state}=[fill=gray!20,draw=black,thick,text=black,scale=0.72]
\node[state,draw=black,scale=1, line width=0.4mm ]         (A)  {$1$};
\node[state,draw=black,scale=0.8]         (B) [ above right of =A] {$2$};
\node[state,draw=black,scale=0.8]         (C) [ right of =B] {$3$};
\node[state,draw=black,scale=0.8]         (C2) [ right of =C] {$4$};
\node[state,draw=black,scale=1]         (D) [below right of =C2] {$\tau +1$};
\path (A) edge [loop left] node {$1$} (A);
\path (A) edge [bend left] node {$1$} (B);
\path (B) edge [bend left] node {$2$} (A);
\path (B) edge [bend left] node {$2$} (C);
\path (C) edge [bend left] node {$2$} (A);
\path (C) edge [bend left] node {$2$} (C2);
\path (C2) edge [bend left] node {$2$} (A);
\path (C2) edge [bend left,dashed] node {} (D); 
\path (D) edge [bend left] node {$2$} (A); 
\end{tikzpicture}
\hspace{0.5cm}
\begin{tikzpicture}[->,>=stealth',shorten >=1pt,auto,node distance=2cm, font=\small, semithick]
\tikzstyle{every state}=[fill=gray!20,draw=black,thick,text=black,scale=0.8]
\node[state,draw=black,scale=1, line width=0.4mm ]         (A)  {$1$};
\path (A) edge [loop left] node {$1$} (A);
\path (A) edge [loop above] node {$\{1,2 \}$} (A);
\path (A) edge [loop right, dashed] node {} (A);
\path (A) edge [loop below] node {$\{1,2,...,2 \}$} (A);
\end{tikzpicture}
\end{center}
\caption{Left, a maximum dwell-time constraints graph $\cG(\cV,\cE)$ for a system consisting  of two modes, $\tau>1$. Right, the Reduced graph taking $\cY:=\{1\}$.}
\label{fig6666}
\end{figure}

In the setting of maximum dwell-time specifications, the system is allowed to
switch between a set of different, possibly unstable, $N$ modes. However,
to remain in a set of modes is allowed only for a limited period, namely for $\tau\geq 1$ time instants at most.
We may express such constraints in a switching constraints graph $\cG(\cV,\cE)$ consisting of a basic, unavoidable node and additional nodes which realize the constraints.
An example of such a system consisting of two modes is in Figure~\ref{fig6666}.
In Proposition~\ref{theorem14}, we show that Proposition~\ref{theorem9}
can be further refined in this case.
Since in the studied setting there is only one unavoidable node, we assign to it the number $1$ without any loss of generality.

\begin{proposition}\label{theorem14}
Consider a system with $N$ modes, subject to maximum dwell-time constraints, $\tau\geq 1$, expressed in the form of the System \eqref{eq_sys1}--\eqref{eq_con3}, and with $\cX_j=\cX\subset\R^n$, for all $j\in\cV$.
Consider the unavoidable set of nodes  $\cY=\{1\}$ and the Reduced
System \eqref{eq_sysred1}--\eqref{eq_sysred3} subject to the constraints \eqref{eq_conred1}, \eqref{eq_conred3}. Let $\tilde{\cS}_M^{1}$ be the maximal invariant set with respect to the Reduced System. Then, the maximal invariant multi-set $\{\cS_M^i \}_{i\in\cV}$ with respect to the System \eqref{eq_sys1}--\eqref{eq_sys3} and the constraints \eqref{eq_con1}, \eqref{eq_con3} is
\begin{equation} \label{eq_th14_1}
    \resizebox{1.02\hsize}{!}{%
$\cS^j_M    :=
\begin{cases}
      \quad \quad \quad \quad \quad \quad \quad \quad \quad \quad \quad \quad \quad \tilde{\cS}_M^1, & j=1, \\
    \left(	\bigcap\limits_{\{i\in m(j,1)\setminus \{j,1\} \}}	\cC(\sigma(j,i), \tilde{\cS}_M^1  )	\right) \cap \cX, &   j\in \cV\setminus \{ 1\}.
    \end{cases}$
    }
  \end{equation}
\end{proposition}

\begin{example}\label{example8}
{\color{black}{ We consider a system under maximum dwell time constraints with $\tau=2$, consisting of two modes. In specific, we consider $A_1=A_2=A\in\R^{2\times 2}$ and two different disturbance sets $\cW_i$, $i=1,2$, where
$A=\left[\begin{smallmatrix}0.7747 & 1.2483 \\ -0.4 & 0.6 \end{smallmatrix}\right]$,
$ \cW_1=\B_{\infty}(10^{-4})$, $\cW_2=250\cW_1.$
\begin{figure}[ht]
\psfrag{x}[][][0.8]{\color{black}{$x_1$}}
\psfrag{y}[][][0.8]{\color{black}{$x_2$}}
\psfrag{Z}[][][0.9]{\color{black}{$\cS_\cV$}}
\psfrag{M}[][][0.9]{\color{black}{$\cX_i$}}
\begin{center}
\includegraphics[height=4.5cm]{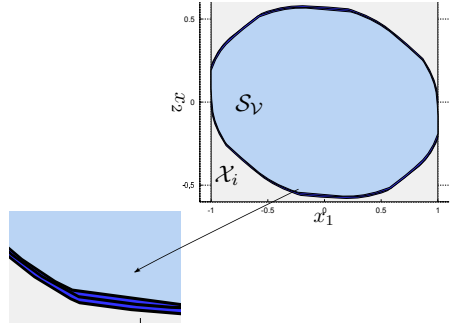}
\end{center}
\caption{ \color{black}{ Example \ref{example8}, the constraint set $\cX_i$ (gray), the maximal invariant multi-set $\{\cS_M^i \}_{i\in[3]}$ (dark blue) and the maximal safe set with respect to all nodes of the graph $\cS_\cV$ (light blue). A zoomed portion of the figure is in lower left.
}}
\label{fex8}
\end{figure}
Such systems where the switching is in the disturbance sets may appear in practical situations, e.g., in networked control systems where channels with different signal to noise ratios are available for communication.
We calculate the maximal invariant multi-set $\{ \cS_M^i\}_{i\in\cV}$, where the maximum dwell-time constraints graph consists of three nodes, i.e., $\cV=\{1,2,3\}$. The system is subject to state constraints $\cX_i=\B_\infty(1)$, $i\in\cV$. First, we compute the maximal invariant set $\tilde{\cS}_M^1$ for the Reduced System, which is reached in $20$ iterations. It is worth noting that by utilizing directly Theorem~\ref{theorem3} we  compute the maximal invariant multi-set in $58>20$ iterations.   Next, utilizing Proposition~\ref{theorem14}, we compute the maximal invariant multi-set $\{ \cS_M^i\}_{i\in\cV}$ for the original system.  In Figure~\ref{fex8}, the  constraint set $\cX_i$ is shown in grey and the elements of the multi-set $\{\cS_M^i \}_{i\in\cV}$ are shown in dark blue. Following Corollary~1, the maximal safe set with respect to all nodes is $\cS_\cV=\cS_M^1\cap\cS_M^2\cap\cS_M^3$ and is shown in Figure~\ref{fex8} in light blue color. 
}}
\end{example}

\subsection{Non-convex approximations of the minimal invariant set for arbitrary switching systems} \label{section_app2}

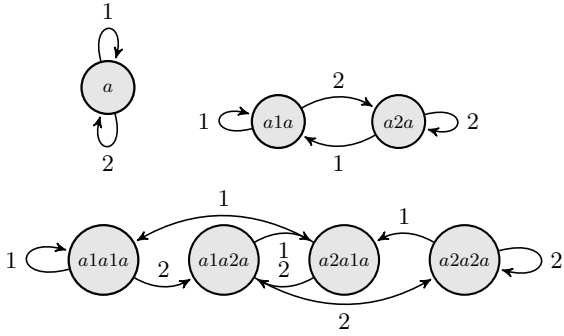
\begin{figure}[h]
\begin{center}
\begin{tikzpicture}[->,>=stealth',shorten >=1pt,auto,node distance=2cm, font=\small, semithick]
\tikzstyle{every state}=[fill=gray!20,draw=black,thick,text=black,scale=0.8]
\node[state,draw=black,scale=1, line width=0.3mm ]         (A)  {$a$};
\path (A) edge [loop above] node {$1$} (A);
\path (A) edge [loop below] node {$2$ } (A);
\end{tikzpicture}
\hspace{0.5cm}
\begin{tikzpicture}[->,>=stealth',shorten >=1pt,auto,node distance=2cm, font=\small, semithick]
\tikzstyle{every state}=[fill=gray!20,draw=black,thick,text=black,scale=0.8]
\node[state,draw=black,scale=1, line width=0.3mm ]         (A)  {$a1a$};
\node[state,draw=black,scale=1]         (B) [ right of =A] {$a2a$};
\path (A) edge [loop left] node {$1$} (A);
\path (A) edge [bend left] node {$2$} (B);
\path (B) edge [bend left] node {$1$} (A);
\path (B) edge [loop right] node {$2$} (B);
\end{tikzpicture}
\hspace{0.5cm}
\begin{tikzpicture}[->,>=stealth',shorten >=1pt,auto,node distance=2cm, font=\small, semithick]
\tikzstyle{every state}=[fill=gray!20,draw=black,thick,text=black,scale=0.8]
\node[state,draw=black,scale=1, line width=0.3mm ]         (A)  {$a1a1a$};
\node[state,draw=black,scale=1]         (B) [ right of =A] {$a1a2a$};
\node[state,draw=black,scale=1]         (C) [ right of =B] {$a2a1a$};
\node[state,draw=black,scale=1]         (D) [ right of =C] {$a2a2a$};
\path (A) edge [loop left] node {$1$} (A);
\path (A) edge [bend right] node {$2$} (B);
\path (B) edge [bend left] node [swap] {$1$} (C);
\path (B) edge [bend right] node [swap] {$2$} (D);
\path (C) edge [bend left] node [swap]  {$2$} (B);
\path (C) edge [bend right] node [swap]  {$1$} (A);
\path (D) edge [loop right] node  {$2$} (D);
\path (D) edge [bend right] node [swap]  {$1$} (C);
\end{tikzpicture}
\end{center}
\caption{Upper left, a graph representing an arbitrary switching system consisting of two modes. Upper right, its $1$-Path-Dependent Lifted Graph. Lower part, its $2$-Path-Dependent Lifted Graph.}
\label{fig66}
\end{figure}

We revisit the problem of computing the minimal invariant set for arbitrary switching
systems. By a modification of the results in  \cite[Section 4.3]{TechReport2005}, we may compute invariant $\epsilon$--approximations of either the minimal invariant set or the minimal convex invariant set. In this subsection, we compute non-convex approximations of the minimal invariant set of a \emph{controlled} complexity by utilizing Theorem~\ref{theorem11}(i). In detail, by applying the $P$-Path-Dependent Lift, we may approximate the minimal invariant set as the union of a finite number of convex sets. For example, for the case of $N=2$ modes, the $1$-Path-Dependent Lifted graph consists of two nodes, as shown in the center of Figure~\ref{fig66}. The $2$-Path-Dependent Lifted graph is shown in the right part of Figure~\ref{fig66}.

{ \color{black}{
\begin{example} \label{example9}
We consider the numerical example in \cite[Example 1]{TechReport2005} that concerned a two--dimensional linear difference inclusion, consisting of two extreme subsystems.  In the context of our study, we consider the system as a linear switching systems that switches arbitrarily between these two modes. We illustrate that the $1$-Path-Dependent Lift allows the exact approximation of the minimal invariant set using convex operations.
To this purpose, by  considering the $1$-Path-Dependent Lift, 
we obtain the Lifted Graph, as shown in the center of Figure~\ref{fig66}. 
First, we use Theorem~\ref{theorem1}(ii) and compute an inner $\epsilon$--approximation of the minimal convex invariant multi-set of the Lifted System for $\epsilon=10^{-3}$. To this purpose, we calculate the pair  $(\Gamma,\rho)=(4.0716,0.7368)$ in \eqref{eq_lemma1_1}, and $\alpha=10$ in the statement of Theorem~\ref{theorem1}(i). In this case, the number of iterations required for the $10^{-3}$--approximation is upper bounded by l=40. The algorithmic implementation in MATLAB converged in $20<40$ iterations, when it reached the machine precision. The multi-set sequence $\{\cS_i^{a1a}, \cS_{i}^{a2a} \}_{i\in\{1,...,20 \}}$ is shown in Figure~\ref{f10}, where the sets $\{\cS_i^{a1a}\}_{i\in\{1,...,20 \}}$ and $\{\cS_i^{a2a}\}_{i\in\{1,...,20 \}}$ are depicted in orange and  blue color respectively. 
Next, by utilizing Theorem~\ref{theorem11}(i) and by taking into account Remark~\ref{remark_cortheorem6}, a non-convex, $\epsilon$--approximation of the minimal invariant set of the original system is $\check{\cS}=\cS_{20}^{a1a}\cup\cS_{20}^{a2a}$. It is worth noting that the set $\check{\cS}$ coincides with the true, non-convex, inner approximation of the minimal invariant set for this particular case.

\begin{figure}[ht]
\psfrag{x}[][][0.8]{\color{black}{$x_1$}}
\psfrag{y}[][][0.8]{\color{black}{$x_2$}}
\psfrag{Z}[][][0.9]{\color{black}{$\{\cS_{i}^{a1a}\}_{i\in[20]}$}}
\psfrag{M}[][][0.9]{\color{black}{$\{\cS_{i}^{a2a}\}_{i\in[20]}$}}
\begin{center}
\includegraphics[height=4cm]{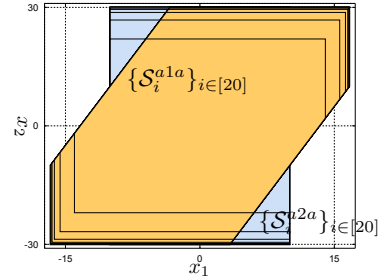}
\end{center}
\caption{ \color{black}{ Example 9, the sets $\{\cS_i^{a1a}\}_{i\in[20]}$ and $\{\cS_i^{a2a}\}_{i\in[20]}$ are depicted in orange and  blue color respectively. The minimal invariant set for the system is $\check{\cS}=\cS_{20}^{a1a}\cup\cS_{20}^{a2a}$.
}}
\label{f10}
\end{figure}
\end{example}
}}

\subsection{Efficient computation of the maximal invariant set for linear systems}
From Theorem~\ref{theorem99}(ii) we can provide an alternative of computing the maximal invariant multi-set of the system
\eqref{eq_sys1}--\eqref{eq_sys3} subject to \eqref{eq_con1}, \eqref{eq_con3}  in two steps. More specifically, one  can compute first the maximal invariant multi-set for the T-product Lifted System \eqref{eq_liftsys1_1}--\eqref{eq_liftsys1_3} subject to \eqref{eq_liftcon1_1}, \eqref{eq_liftcon1_2} and consequently utilize \eqref{eq_th10_1}
of Theorem~\ref{theorem99}.
Let us consider the linear system
\begin{equation} \label{eq_linear}
x(t+1)=Ax(t)
\end{equation}
where $A\in\R^{n\times n}$,  subject to the constraints $x(t)\in\cX, \quad \forall t\geq 0$,
$\cX\subset \R^n$.
\begin{definition}\label{definition_basicoperation}
Given a set $\cS\subset\R^n$ and a matrix $A\in\R^n$, we define as a \emph{basic iteration}
the set mapping $f_-(\cS)  :=\{x: Ax \in \cS \}$.
\end{definition}
  The \emph{basic iteration} of Definition~\ref{definition_basicoperation} corresponds to the one-step backward reachability map of the linear System \eqref{eq_linear}.
The following Proposition suggests that we can utilize the $T$-product Lifted System for linear systems and compute the maximal invariant set
 in a number of basic iterations  \emph{proportional to the square root of the number of iterations required using the classical approach}, e.g., \cite{BlanMi:08}, which is a special case of   Theorem~\ref{theorem3} for the limiting case of a constraint graph $\cG(\cV,\cE)$ with $\cV=\{ 1\}$, $N=1$, $\cE=\{(1,1,1) \}$.

\begin{proposition} \label{proposition7}
Under Assumptions~\ref{ass1}, \ref{ass3}, consider the System \eqref{eq_linear}
 and a pair $(\Gamma,\rho)$, $\Gamma\geq 1$, $\rho\in (0,1)$ satisfying \eqref{eq_exprel}. Let $R:= \max\{R: \B(R)\subseteq \cX 	 	 \}$,
 $c:=\min \{ c: \B(c)\supseteq \cX\}$. Let
$k=\left\lceil \log_{\rho}\left(\frac{R}{\Gamma c}\right)\right\rceil$.
Then, the maximal invariant set can be computed after $k^\star$ basic iterations, where
\begin{equation}\label{eq_proposition7_1}
k^\star=\left\lceil 2\sqrt{k}-1 \right\rceil.
\end{equation}
\end{proposition}
\begin{remark}
The result \eqref{eq_proposition7_1} follows by finding the optimal lift $T\geq 1$ which minimizes the total number of iterations required to compute the maximal invariant set of  the lifted system and to transform to the maximal invariant set for the original system. Since we are dealing with linear systems, the scalar $\rho$ corresponds to the spectral radius of the matrix $A$, which may be retrieved exactly by an eigenvalue analysis or by solving the related Lyapunov inequality.
\end{remark}

\begin{remark}
The same reasoning may carry on after choosing specific families of constraint sets e.g., polyhedral or semi-algebraic sets, or by considering the more general case of constrained switching systems. In detail, as in the setting of  \cite[Section V, Lemma 4]{Bemporad:11}, we can  obtain an explicit bound on the number of linear inequalities in case the constraint set $\cX$ is a polyhedral set. 
\end{remark}

\begin{example}\label{example10}
We study the triple integrator $\dot{x}(t)=Ax(t)+Bu(t)$,  $A= \left[\begin{smallmatrix}
0 & 1 & 0 \\ 0 & 0 & 1\\ 0 & 0 & 0
\end{smallmatrix}\right], B=\left[\begin{smallmatrix} 0\\ 0 \\ 1\end{smallmatrix}\right]$. In specific,
we consider its discretized version with $\tau=0.3$, i.e., $x(k+1)=A_dx(k)+B_du(k)$, with $A_d=I+\tau A$, $B_d=\tau B$.
We consider a linear state feedback controller which is the solution of the LQR problem setting $Q=1$, $R=10^7$, i.e., $u(k)=-Kx(k)=-\left[\begin{smallmatrix} 3\cdot 10^{-4} & 9.2\cdot 10^{-3} & 0.1363\end{smallmatrix}\right]x(k)$.
Moreover, we consider the constraint set $\cX=\{ x\in\R^3: |x|\leq w\}$, with $w=\left[\begin{smallmatrix} 4 & 3 & 1 \end{smallmatrix}\right]^\top$. We compute the maximal admissible invariant set for the closed-loop system. By utilizing Theorem~\ref{theorem3} we compute the theoretical upper bound in the number of iterations in the backward reachability algorithm\footnote{\color{black}{We compute a quadratic Lyapunov function that corresponds to the contractive set with contraction factor equal to the spectral radius of the closed-loop system $\rho(A_d+B_dK)=0.9898$. We also compute $\Gamma=286$, $R=1$, $c=4$.} } to be $k=689$. The maximal admissible invariant set $\cS_M$ is computed in MATLAB in $164<684$ iterations.
Following Proposition~\ref{proposition7}, we compute the maximal admissible invariant set $\cS_M$ in two steps. We first compute the  T-lifted System which will minimize the required number of basic iterations, i.e., we set $T=\sqrt{k}=27$. The number of iterations for computing the maximal admissible invariant set $\check{\cS}_M$ for the lifted system is upper bounded by  $k^\star=27$ steps while it is computed in MATLAB in $7$ iterations. Following Theorem~\ref{theorem99}, the transformation to the maximal admissible invariant set of the original system requires $T-1=26$ additional basic iterations. Concluding,  by applying Proposition \ref{proposition7} the maximal invariant set is retrieved in overall $7+26=33$ iterations, while the theoretical upper bound is $52$ basic iterations. In Figures~\ref{fig:test1} and \ref{fig:test2} the maximal admissible invariant sets $\cS_M$ and $\check{\cS}_M$ for the system and the lifted system are shown respectively. 
\begin{figure}[ht] 
\begin{center} 
\includegraphics[height=3.5cm]{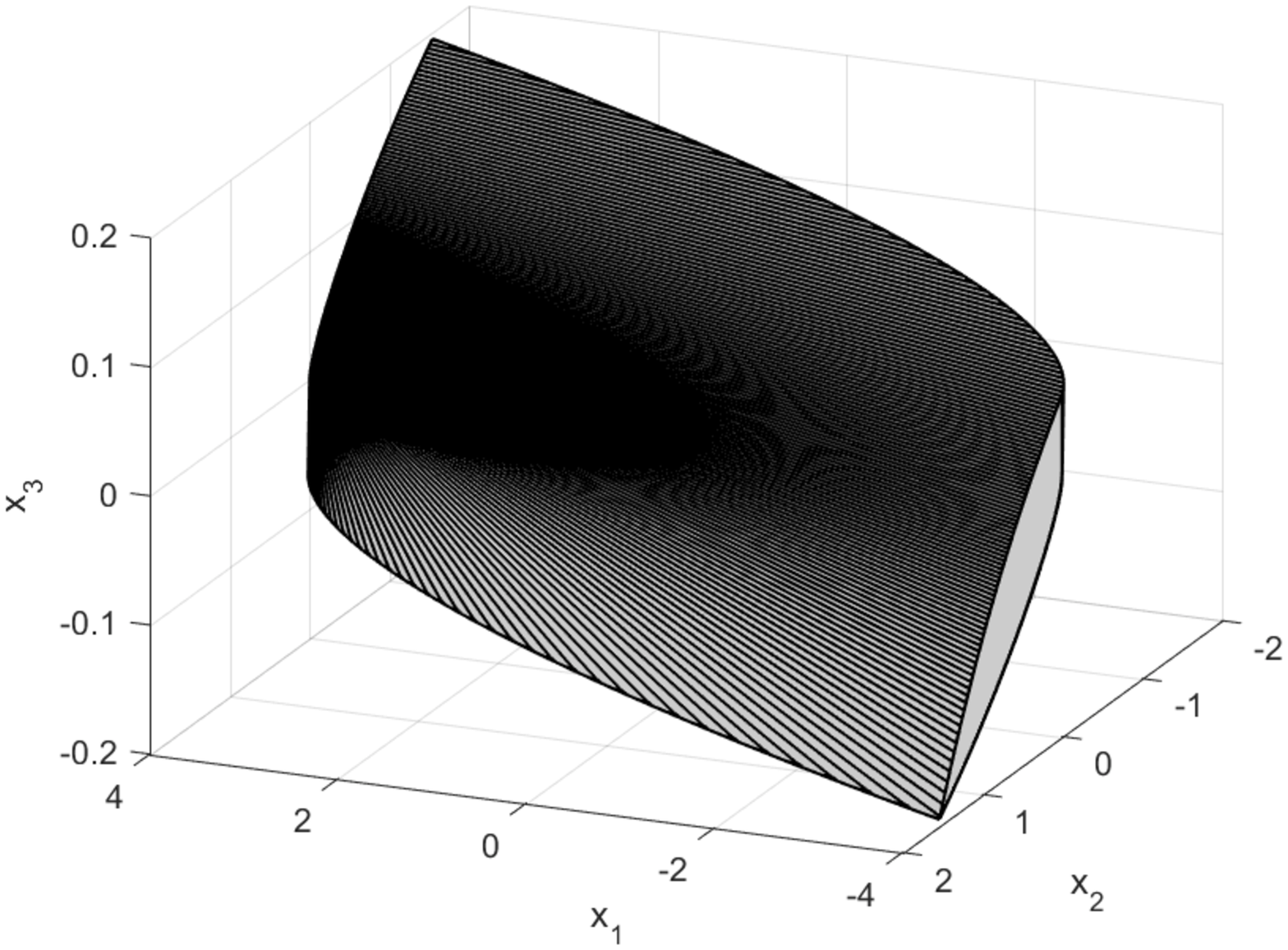}
\end{center}
\caption{ \color{black}{ Example~\ref{example10}, the set $\cS_M$.
}}
\label{fig:test1}
\end{figure}
\begin{figure}[ht]
\begin{center}
\includegraphics[height=3.5cm]{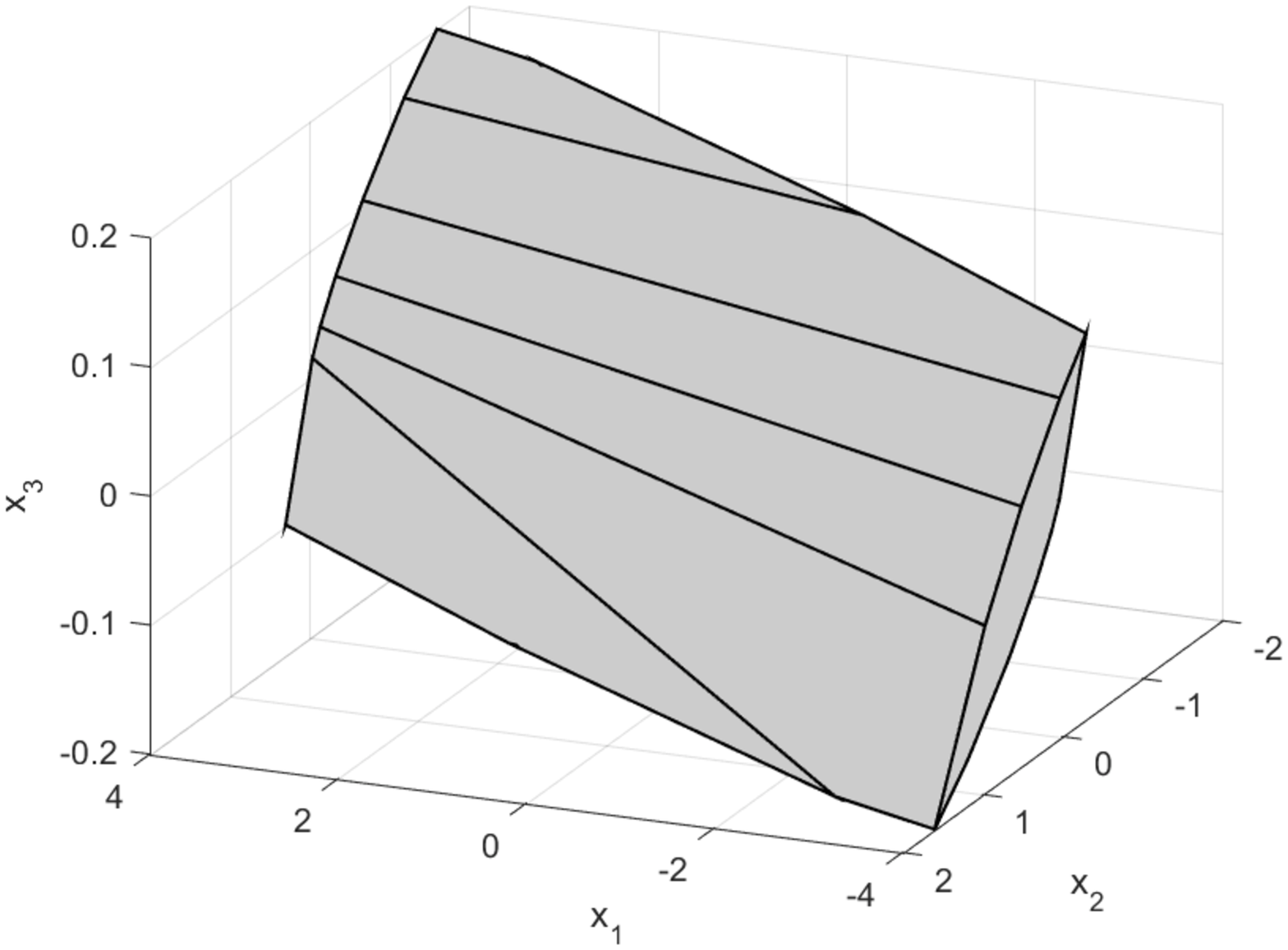}
\end{center}
\caption{ \color{black}{ Example~\ref{example10}, the set $\check{\cS}_M$ for the T-lifted system, $T=27$. }}
\label{fig:test2}
\end{figure}
\end{example}

\section{Conclusions}
The iterative computation of invariant sets has attracted a lot of attention in recent years. This is partly due to many situations in modern engineering, where safety-critical, ressource-aware, embedded, or Cyber-Physical constraints can be tackled by such concepts. A particular effort has been devoted to switching systems, because of their important modelling power. The question at the basis of this paper was:
how do existing techniques generalize when the system is not switching arbitrarily, but has its switching signal constrained by an automaton? 
We have shown that invariant sets need to be generalized in this new setting, and we have exploited the concept of multi-set to this purpose. With this tool in hand, we have developed generalizations of the algorithms previously existing for arbitrarily switching systems. We have addressed their computational complexity, and reduced it by exploiting combinatorial constructions from the automata theory. As a proof-of-concept, we addressed several practical applications in control and showed that significant ameliorations are possible, either in terms of improving computational efficiency or by further refining the notions of invariance.

\appendix

\section{Proofs}
\subsection{Proof of Fact~\ref{factt1}.}
When $l=0$, the relation \eqref{eq_fact1_1} holds trivially. We assume that \eqref{eq_fact1_1} holds for $l=k$.
Then, $\cF_{k+1}^j  = \bigcup\limits_{(s,j,\sigma)\in\cE}\cR(\sigma,\cF_k^s)=\bigcup\limits_{(s,j,\sigma)\in\cE}\cR\left(\sigma, \bigcup\limits_{i\in\{0,...,k\}}	\cF_i^s\right)= \bigcup\limits_{(s,j,\sigma)\in\cE}\bigcup\limits_{i\in\{0,...,k\}}\cR\left(\sigma, 	\cF_i^s\right) =  \bigcup\limits_{i\in\{0,...,k \}} \cF_{i+1}^j =$ 

\noindent $\bigcup\limits_{i\in\{0,...,k \}} \cF_{i+1}^j\cup \{0\}= \bigcup\limits_{i\in\{0,...,k+1\}} \cF_{i}^j,$
thus, \eqref{eq_fact1_1} holds for all $l\geq 0$. $\blacksquare$

\subsection{Proof of Proposition~\ref{proposition1}.}
The left inclusion holds from Fact~\ref{factt1}.
 To prove the right inclusion,
from Assumption~\ref{ass3} there exist scalars $\Gamma$, $\rho\in(0,1)$ such that \eqref{eq_lemma1_1} holds.
Setting $\cZ_l:=\rho^l\Gamma  \cW^\star$ we have
the relations in the top of page 17,
\begin{figure*}[t]
 \begin{align}
 \cF_{l+1}^j & =  \bigcup\limits_{(s_l,j,\sigma_l)\in\cE} \cR( \sigma_l, \cR(   \ldots		(	\sigma_1, \bigcup\limits_{(s_1,s_2,\sigma_1)\in\cE}\cR(\sigma_0,   \bigcup\limits_{(s_0,s_1,\sigma_0)\in\cE}\cF_0^{s_0}   )			)	\ldots	) 				 )  \nonumber \\
 & =  \bigcup\limits_{(s_l,j,\sigma_l)\in\cE}A_{\sigma_l}(...( \bigcup\limits_{(s_{1},s_2,\sigma_{1})\in\cE}A_{\sigma_1}( 
  \bigcup\limits_{(s_0,s_1,\sigma_0)\in\cE}(A_{\sigma_0}\cF_0^{s_0}\oplus\cW_{\sigma_0})    )  \oplus \cW_{\sigma_1}...))\oplus\cW_{\sigma_l} \nonumber \\  
&  = \bigcup\limits_{(s_l,j,\sigma_l)\in\cE}(	...(\bigcup_{(s_1,s_2,\sigma_1)\in\cE}A_{\sigma_l}...A_{\sigma_1}\cN_0^{s_1}\oplus	  A_{\sigma_l}...A_2\cW_{\sigma_1} )...) \oplus A_{\sigma_l}\cW_{\sigma_{l-1}} \oplus\cW_{\sigma_l} \label{eq_al3} \\
& \subseteq \bigcup\limits_{(s_l,j,\sigma_l)\in\cE}(	...(	\bigcup_{(s_1,s_2,\sigma_1)\in\cE}\cZ_l\oplus A_{\sigma_l}...A_2\cW_{\sigma_1}		)
  ...) \oplus A_{\sigma_l}\cW_{\sigma_{l-1}} \oplus\cW_{\sigma_l} = \cF_l^j\oplus\cZ_l. \label{eq_al4}
\end{align}
\end{figure*}
where  in \eqref{eq_al3} and \eqref{eq_al4} we use the relations \eqref{eq_setseq1_1} and \eqref{eq_lemma1_1} correspondingly. Thus, the right inclusion in \eqref{eq_proposiition1_1} holds. $\blacksquare$

\subsection{Proof of Theorem~\ref{theorem1}}
(i) From Proposition~\ref{proposition1}, the set sequence $\{ \cF_i^j \}_{i\geq 0}$, for each $j\in\cV$, is monotonically non-decreasing and is a Cauchy sequence. Thus, the set sequence is convergent in the space of compact sets having as metric the Hausdorff distance and
 a limit $\cF_\infty^j$ exists, for all $j\in\cV$.
(ii) The left inclusion holds from Fact~\ref{factt1}. From Proposition~\ref{proposition1},  it holds that $\hausd{\cF_l^j}{\cF_{l+1}^j}\leq \Gamma\alpha\rho^l$, for any $l\geq 0$, $j\in\cV$. Consequently,
for any $j\in\cV$, $m\geq 1$, $l\geq 0$ we have $\cF_{l+m}^j\subseteq \Gamma \alpha \rho^l \frac{1-\rho^m}{1-\rho}\B(1)\oplus\cF_l^j$.
Taking the limit as $m\rightarrow \infty$, it follows that $\cF_{\infty}^j\subseteq \frac{\Gamma\alpha\rho^l}{1-\rho}\B(1)\oplus\cF_l$.
Thus, relation \eqref{eq_theorem1_1} is satisfied for any $l\geq\lceil \log_{\rho} ( \frac{\epsilon(1-\rho)}{\alpha \Gamma} ) \rceil$.
 (iii) Invariance of the multi-set $\{\cF_\infty^j\}_{j\in\cV}$ follows directly from Fact~\ref{factt1}. To show minimality, we use a similar reasoning as in \cite[Lemma 3.1]{BlanchMianiSznaier:97}: Let us assume there exists a compact invariant multi-set $\{\cS^j\}_{j\in\cV}$ and an index $j^\star\in\{1,...,M \}$  such that  $\cF_{\infty}^{j^\star}\nsubseteq \cS^{j^\star}$.
Then, for any $x(0)\in\R^n$, for any $z(0)\in\cV$ and under Assumptions~\ref{ass2}, \ref{ass4},  we pick $w(t)=0$, for all $t\geq 0$. We choose a solution $(x(t),z(t))$,
 $t\geq 0$, for which there exists a time sequence $\{t_i\}_{i\geq 0}$ such that $z(t_i)=j^\star$, $i\geq 0$. 
From Assumption~\ref{ass1}, $x(t)\rightarrow 0$ as $t\rightarrow \infty$. Since $\cS^{j^\star}$ is compact, $x(t_i)\in\cS^{j^\star}$ and $\{x(t_i)\}_{i\geq 0}$ converges to $0$,  
it necessarily holds that $0\in\cS^{j^\star}$.
However, since $\cS^{j^\star}$ is a member of the invariant multi-set $\{\cS^j\}_{j\in\cV}$,
we have  from Fact~\ref{factt1} that $\cF_{\infty}^{j}\subseteq \cS^{j}$ for all $j\in\cV$, which is a contradiction. Thus, $\cF_{\infty}^{j^\star}\subseteq\cS^{j^\star}$
and $\{\cF_{\infty}^j \}_{j\in\cV}$ is the minimal invariant multi-set. $\blacksquare$

\subsection{Proof of Proposition~\ref{theorem2}}
(i) For $l=0$ the relation holds. Similarly to \cite[Section 3]{TechReport2005}, \cite[Proposition 1]{Nikos14}, we assume that it holds for $l=k$. For $l=k+1$ it follows that
$\conv{\cF_{l+1}^j} = \conv{\bigcup\limits_{(s,j,\sigma)\in\cE}\cR(\sigma,\cF_l^s)}  = \conv{\bigcup\limits_{(s,j,\sigma)\in\cE}\conv{\cR(\sigma,\conv{\cF_l^s})}}=$

$ = \conv{\bigcup\limits_{(s,j,\sigma)\in\cE}\cR_\text{C}(\sigma,\s{\cF}_l^s)}= \conv{\s{\cF}_{l+1}^j}.$
Statements (ii)--(iv) can be proved using the same arguments as in Theorem~\ref{theorem1}, by reproducing the results in Fact~\ref{factt1}, Proposition~\ref{proposition1} for the multi-set sequence \eqref{eq_setseq_min2_1}, \eqref{eq_setseq_min2_2}. In the right inclusion of (iv), we use the fact that the convex hull and the Minkowski sum operators commute.
$\blacksquare$

\subsection{Proof of Theorem~\ref{theorem22}}
\noindent (i)  
For an arbitrary edge $(i,j,\s{\sigma})\in\cE$ we have
$A_{\s{\sigma}}\cD_k^i\oplus \cW_{\s{\sigma}}  = \frac{1}{1-\lambda} A_{\s{\sigma}} [  \bigcup\limits_{ (s_{k-1},i,\sigma_{k-1})\in\cE } A_{\sigma_{k-1}}\cF_{k-2}^{s_{k-1}}\oplus \cW_{\sigma_{k-1}} ]\oplus\cW_{\s{\sigma}}=\cdots$
 $= \frac{1}{1-\lambda}A_{\s{\sigma}} (   \bigcup\limits_{\substack{ \{s: |\sigma(s,i)|=k-1,  \\  m(s,i)=\{\sigma_{k-1},...,\sigma_1 \} \} }  }
		\bigoplus\limits_{q=1}^{k-1}  \prod\limits_{r=1}^{k-q} A_{\sigma_{k-r}}\cW_{\sigma_q}		  )\oplus \cW_{\s{\sigma}}.$
For any switching sequence $m(s,j)=\{ \s{\sigma},\sigma_{k-1},...,\sigma_{2},\sigma_1  \}$
of length $k$, i.e., $|m(s,j)|=k$, by hypothesis we have $A_{\s{\sigma}}A_{\sigma_{k-1}}....A_{\sigma_2}\cW_{\sigma_1}\subseteq \lambda \cN_{\cap}^j$,
and since $\cN_{\cap}^j\subseteq \cW_{\sigma}$, it follows that $A_{\s{\sigma}}A_{\sigma_{k-1}}...A_{\sigma_2}\cW_{\sigma_1}\subseteq \lambda \cW_{\s{\sigma}}$.
Thus,
$A_{\s{\sigma}}\cD_k^j\oplus\cW_{\s{\sigma}}  \subseteq  \frac{1}{1-\lambda} 
( 	\bigcup\limits_{{\substack{ \{s: |\sigma(s,i)|=k-1,  \\  m(s,i)=\{\sigma_{k-1},...,\sigma_2 \} \} }  } }	  		
\bigoplus\limits_{q=2}^{k-1} \prod_{r=1}^{k-q} A_{\sigma_{k-r}}\cW_{\sigma_q} 	)$  $\oplus
\left(  \frac{\lambda}{1-\lambda} + 1		 \right)\cW_{\s{\sigma}}$
$\subseteq \frac{1}{1-\lambda} \cF_{k-1}^j =\cD_{k}^j,$
%
and by Proposition~\ref{prop_altdef} the multi-set $\{ \cD_{k}^j \}_{j\in\cV}$
 is invariant with respect to the System.  (ii) The left inclusion holds by definition since from (i) the multi-set $\{\cD_{k}^j\}_{j\in\cV}$ is invariant. Given $\epsilon>0$, we pick $k\geq 1,\lambda\in(0,1)$ such that  \eqref{eq_prop2_1} holds and $\frac{\lambda}{1-\lambda} \cF_{k-1}^j\subseteq \B(\epsilon)$, for all $j\in\cV$.
There always exists such a pair $(k,\lambda)$ since $\lambda$ can be made arbitrarily small and by Fact~\ref{factt1} and Theorem~\ref{theorem1}  the set $\cF_{k-1}^j\subseteq \cF_\infty^j= \cS_m^j$
is bounded.
Consequently, we have that
$
 \cD_k^j =  \left(1 + \frac{\lambda}{1-\lambda} \right)\cF_{k-1}^j\subseteq\cF_{k-1}^j\oplus \B(\epsilon)\subseteq\cF_{\infty}^j\oplus \B(\epsilon)= \cS_{m}^j\oplus \B(\epsilon). \blacksquare
$

\subsection{Proof of Lemma~\ref{fact3}}

For any $t\geq 0$, $x(0)\in\R^n$, $\{z(t)\}_{t\geq 0}$ and any $\sigma(t)$, $t\geq 0$, satisfying \eqref{eq_con1} we have
$x(t)=x_1(t)+x_2(t)$, where $x_1(t)  :=\prod_{i=0}^{t-1}A_{\sigma(t-1-i)}x(0)$ and  $x_2(k)  := \sum_{j=0}^{t-2}\left(\prod_{i=0}^{t-2-j}A_{\sigma(t-1-i)}w(j)\right)+w(t-1).$
Under Assumption~\ref{ass2}, from \eqref{eq_exprel} $\Gamma\geq 1$, $\rho\in[0,1)$ such that $x_1(t)\in\Gamma\rho^t c\B(1)$. Moreover, by definition, $x_2(t)\in\cF_t^{z(t)}$, where $\{\cF_i^j\}_{j\in\cV}$, $i\geq 0$, generated by \eqref{eq_setseq_min1_1}, \eqref{eq_setseq_min1_2}. Thus, $x_2(t)\in\cS_m^{z(t)}$, for all $t\geq 0$.
Consequently,
$x(t)\in \left(\Gamma c\rho^t\B(1)\oplus \cS_m^{z(t)}\right),
$ or,
$d(x(t),\cS_m^{z(t)})\leq \Gamma c \rho^t$.
Thus,  $d(x(t),\cS_{m}^{z(t)})\leq \epsilon$, for any $t\geq l^\star$, for any $l^\star\geq \left\lceil \log_{\rho} \left( \frac{\epsilon}{c \Gamma} \right)  \right\rceil$.
$\blacksquare$

\subsection{Proof of Theorem~\ref{theorem3}}
(i) Using the same reasoning as in the proof of Lemma~\ref{fact3}, for any initial condition $(x(0),z(0))$,  $x(0)\in\cX_z(0)$, $z(0)\in\cV$ it holds that
$\|x(t) \|\leq \Gamma \rho^t c +r_{z(t)}$, for all $t\geq 0$.
Consequently, we have that $\|x(t) \|\leq R_{z(t)}$, or, equivalently, $x(t)\in\cX_{z(t)}$, for all $t\geq \s{k}$, where $\s{k}$ is given in \eqref{eq_th5_1}.
Let us assume that $x(0)\in\cB_{\s{k}}^{z(0)}$ but $x(0)\notin\cB_{\s{k}+1}^{z(0)}$. Then, $x(\s{k}+1)\notin \cX_{z(\s{k}+1)}$ which is a contradiction.
Thus, $\cB_{\s{k}+1}^{z(0)}\supseteq \cB_{\s{k}}^{z(0)}$. Taking into account that $\cB_{l+1}^j\subseteq\cB_l^j$ holds by construction for all $j\in\cV$, $l\geq 0$, the result follows.
(ii) We take similar steps as in the proofs of results concerning the linear case or the case of arbitrary switching, e.g., \cite{Kolmanovsky98}:
From (i) and Proposition~\ref{prop_altdef}(iii), it follows that $\{\cB_{\s{k}}^j\}_{j\in\cV}$ is an admissible invariant multi-set. Suppose that there exists an admissible invariant multi-set $\{\cM^j\}_{j\in\cV}$
and an index $j^\star$ for which 	$\cM^{j^\star}\nsubseteq \cB_{\s{k}}^{j^\star}$. Then, for all $x(0)\in\cM^{j^\star}\setminus \cB_{\s{k}}^{j^\star}$, $z(0)=j^\star$, it follows
that $x(\s{k})\notin\cX_{z(\s{k})}$ and $\{\cM^{j^\star}\}_{j\in\{1,...,M \}}$ is not admissible, which is a contradiction. Thus, $\cM^{j^\star}\subseteq \cB_{\s{k}}^{j^\star}$ and $\{\cB_{\s{k}}^{j}\}_{j\in\cV}$ is the maximal admissible invariant multi-set with respect to the System \eqref{eq_sys1}-\eqref{eq_sys3} and the constraints \eqref{eq_con1}, \eqref{eq_con3}. $\blacksquare$

\subsection{Proof of Lemma~\ref{lemma4}}
We first prove the left inclusion. For $l=0$, it holds $\cF_0^j=\tilde{\cF}_0^j=0$. Assuming the inclusion holds for $l=k\geq 1$, we have for $l=k+1$,  $j\in\cY$, that
\begin{align} 
\tilde{\cF}_{k+1}^j & = \bigcup\limits_{(i,j,\sigma)\in\tilde{\cE}}\cR(\sigma,\tilde{\cF}_k^i)\supseteq \bigcup\limits_{(i,j,\sigma)\in\tilde{\cE}}\cR(\sigma,\cF_{k \theta_m}^i)\nonumber \\
&=\bigcup\limits_{\{ i\in\cY: m(i,j)\cap\cY= \{i,j\} \}}\cR(\sigma(i,j),\cF_{k\theta_m}^i). \label{eq_lem4_2}
\end{align}
On the other hand, we have
\begin{align}
&\cF_{(k+1)\theta_m}^j =  \bigcup\limits_{\substack{\{i\in\cY: |\sigma(i,j)|=\theta_m  \} \\ \cup \{i\in \cV\setminus\cY: |\sigma(i,j)|=\theta_m  \}}}\cR(\sigma(i,j),\cF_{k\theta_m}^i)=...= \nonumber   
\end{align}
\begin{align}
   & =  \bigcup\limits_{p=0}^{\theta_M-\theta_m}\bigcup\limits_{\{ i\in\cY: |\sigma(i,j)|=\theta_m + p \}} \cR(\sigma(i,j),\cF_{k\theta_m -p}^i)\nonumber \\
   &\subseteq \bigcup\limits_{p=0}^{\theta_M-\theta_m}\bigcup\limits_{\{ i\in\cY: |\sigma(i,j)|=\theta_m + p \}} \cR(\sigma(i,j),\cF_{k\theta_m }^i). \label{eq_lem4_3}
\end{align}
In \eqref{eq_lem4_3} we use  Fact~\ref{factt1} and the fact that there cannot be a walk between any two nodes in $\cV$ which is longer than  $\theta_M$ and does not contain at least two nodes in $\cY$.
By merging \eqref{eq_lem4_2} and \eqref{eq_lem4_3} it holds that  $\tilde{\cF}_{k+1}^j\supseteq \cF_{(k+1)\theta_m}^j$, for all $j\in\cY$, thus, the left inclusion in \eqref{eq_lem4_1} holds for all $j\in\cV$, for all $l\geq 0$.
 We use induction to prove the right inclusion in \eqref{eq_lem4_1} as well. To this purpose, for $l=0$, it follows that  $\cF_0^j=\tilde{\cF}_0^j=0$. Assuming it holds for $l=k\geq 1$,
we have
\begin{align}
& \cF_{(k+1)\theta_M}^j  = \bigcup\limits_{ \{ i\in\cV: |\sigma(i,j)|=\theta_M \}}\cR(\sigma(i,j),\cF_{k\theta_M}^i) = \nonumber \\
    & =  \bigcup\limits_{p=0}^{\theta_M-\theta_m}\bigcup\limits_{\{ i\in\cY: |\sigma(i,j)|=\theta_M - p \}} \cR(\sigma(i,j),\cF_{k\theta_M +p}^i) \nonumber \\
     & \supseteq  \bigcup\limits_{p=0}^{\theta_M-\theta_m}\bigcup\limits_{\{ i\in\cY: |\sigma(i,j)|=\theta_M - p \}} \cR(\sigma(i,j),\cF_{k\theta_M }^i)\nonumber  \\
& = \bigcup\limits_{\{ i\in\cY: m(i,j)\cap \cY=\{i,j \} \}}      \cR(\sigma(i,j),\cF_{k\theta_M }^i) = \tilde{\cF}_{k+1}^j, \nonumber
     \end{align}
thus, $\tilde{\cF}_{k+1}^j\subseteq\cF_{(k+1)\theta_M}^j $ for all $j\in\cY$, thus, the right inclusion in \eqref{eq_lem4_1} holds for all $j\in\cY$, for all $l\geq 0$. $\blacksquare$

\subsection{Proof of Theorem~\ref{theorem6}}
Taking the limit in \eqref{eq_lem4_1} as $\l\rightarrow \infty$, we have
from Theorem~\ref{theorem1}(iii) that $\cS_m^j\subseteq \tilde{\cS}_m^j\subseteq \cS_m^j$, thus, $\cS_m^j=\tilde{\cS}_m^j$, for all $j\in\cY$.
Using a similar reasoning as in Lemma~\ref{lemma4}, for all $v\in\cV\setminus \cY$ we have for any $l\geq \theta_M$
$\cF_l^v=\bigcup_{(i,v,\sigma)\in\cE}\cR(\sigma,\cF_{l-1}^i)=\bigcup\limits_{p=1}^{\theta_M}\bigcup\limits_{\{i\in\cY: |\sigma(i,v)|=p  \}}\cR(\sigma(i,v),\cF_{l-p}^i),
$
and taking the limit as $l\rightarrow\infty$ the result follows.
$\blacksquare$

\subsection{Proof of Fact~\ref{fact2}}
For $l=1$ it holds that $\cR(\sigma,\cS_1\oplus\cS_2)=A_{\sigma}(\cS_1\oplus\cS_2)\oplus\cW_\sigma=(A_{\sigma}\cS_1\oplus\cW_\sigma)\oplus A_{\sigma}\cS_2=\cR(\sigma,\cS_1)\oplus\cR_{\text{N}}(\sigma,\cS_2)$.
Suppose that \eqref{eq_fact2_1} holds for $l$. Then,
$\cR(\{\sigma_{i}\}_{i\in[l+1]},\cS_1\oplus\cS_2) = \cR(\sigma_{l+1},   \cR( \{ \sigma_i \}_{i\in[l]}, \cS_{1}\oplus\cS_2  ) )=  \cR(\sigma_{l+1},   \cR( \{ \sigma_i \}_{i\in[l]}, \cS_{1} )\oplus \cR_{\text{N}}(
\{\sigma_i  \}_{i\in[l]},\cS_2) )
 = \cR(\{\sigma_i  \}_{i\in[l+1]},\cS_1)\oplus\cR_{\text{N}}(\{\sigma_i  \}_{i\in[l+1]},\cS_2).$
Thus, \eqref{eq_fact2_1} holds for all $l\geq 1$. $\blacksquare$

\subsection{Proof of Proposition~\ref{theorem7}}
From Theorem~\ref{theorem6} and Theorem~\ref{theorem1}(ii), it follows that
$\tilde{\cF}_l^j\subseteq \tilde{\cS}_m^j=\cS_m^j\subseteq \tilde{\cF}_l^j\oplus\B(\epsilon),  \forall j\in\cY,
$
for any $l\geq \left\lceil \log_{\tilde{\rho}} \left( \frac{\epsilon(1-\tilde{\rho})}{\tilde{\alpha} \tilde{\Gamma} }\right)  \right\rceil.$
Moreover, for any $j\in\cV\setminus \cY$, and for any integer $m\geq 0$, $l\geq \theta_M$ we have
$\tilde{\cF}_{l+m+1}^j $ $ = \bigcup\limits_{p=1}^{\theta_M} \bigcup\limits_{\{ i\in\cY: |\sigma(i,j)|=p \}}\cR(\sigma(i,j),\tilde{\cF}_{l+m}^i)\subseteq
\bigcup\limits_{p=1}^{\theta_M}$ $ \bigcup\limits_{\{ i\in\cY: |\sigma(i,j)|=p \}}\cR(\sigma(i,j),\tilde{\cF}_{l+m-1}^i\oplus \tilde{\Gamma}\tilde{\rho}^{l+m-1}\tilde{\alpha}\B(1))$
$\subseteq \bigcup\limits_{i=1}^{\theta_M} \bigcup\limits_{\{ i\in\cY: |\sigma(i,j)|=p \}}\cR \left(  \sigma(i,j),\tilde{\cF}_{l}^i\oplus 		\left(       \frac{  \tilde{\Gamma}\tilde{\rho}^l\tilde{\alpha}(1-\tilde{\rho}^m)	}{1-\tilde{\rho}}     \right)	\B(1)\right).$
%
Setting $\delta=	\left(       \frac{  \tilde{\Gamma}\tilde{\rho}^l\tilde{\alpha}	}{1-\tilde{\rho}}     \right)$, taking the limit as $m\rightarrow \infty$ and by using Fact~\ref{fact2},
we have for all $j\in\cV\setminus\cY$
$\tilde{\cF}_{\infty}^j    =   \bigcup\limits_{p=1}^{\theta_M}\bigcup\limits_{\{ i\in\cY:|\sigma(i,j)|=p  \}} [ 	\cR(\sigma(i,j),\tilde{\cF}_l^i)\oplus \cR_{\text{N}}(\sigma(i,j),\B(\delta))	]$
$ \subseteq \bigcup\limits_{p=1}^{\theta_M}\bigcup\limits_{\{ i\in\cY:|\sigma(i,j)|=p  \}} [ 	\cR(\sigma(i,j),\tilde{\cF}_l^i)\oplus \Gamma\rho^{|\sigma(i,j)|}\B(\delta)		]$
$\subseteq ( \bigcup\limits_{p=1}^{\theta_M}\bigcup\limits_{\{ i\in\cY:|\sigma(i,j)|=p  \}} 	\cR(\sigma(i,j),\tilde{\cF}_l^i))\oplus \Gamma\rho\B(\delta)	= \tilde{\cF}_l^j\oplus \Gamma\rho\B(\delta),$
which in turn implies relation \eqref{eq_th7_2} for any $l\geq \theta_M$ satisfying
$l\geq \left\lceil \log_{\tilde{\rho}} \left( \frac{\epsilon(1-\tilde{\rho})}{\tilde{\alpha} \tilde{\Gamma}\Gamma \rho  }\right)  \right\rceil.
$
Combining the two inequalities on $l$ the result follows.    $\blacksquare$

\subsection{Proof of Proposition~\ref{theorem8}}
Invariance of $\{\tilde{\cD}_k^j \}_{j\in\cV}$ follows directly by the definition of the mapping $f(\cdot)$ \eqref{eq_fmap} and by applying Theorem~\ref{theorem22} to the Reduced System \eqref{eq_sysred1}--\eqref{eq_sysred3}.
For any $j\in\cY$, under hypotheses and from Theorem~\ref{theorem22} and Proposition~\ref{theorem6} we have that $\cS_m^j=\tilde{\cS}_m^j\subseteq \tilde{\cD}_k^j\subseteq \tilde{\cS}_m^j\oplus \B(\epsilon)=\cS_m^j\oplus \B(\epsilon)$.
For $j\in\cV\setminus \cY$ and by utilizing  Fact~\ref{fact2} and  Theorem~\ref{theorem1} we have 
\begin{align*}
&\tilde{\cD}_k^{j} =\bigcup\limits_{p=1}^{\theta_M}\bigcup\limits_{\{ i\in\cY: |\sigma(i,j)|=p  \}}\cR(\sigma(i,j),\tilde{\cD}_k^j)\\
& = \bigcup\limits_{p=1}^{\theta_M} \bigcup\limits_{\{ i\in\cY: |\sigma(i,j)|=p \}}
\cR(\sigma(i,j), (1+\frac{\lambda}{1-\lambda})\tilde{\cF}_{k-1}^j) 
\end{align*}
\begin{align*}
& \subseteq \bigcup\limits_{p=1}^{\theta_M} \bigcup\limits_{\{ i\in\cY: |\sigma(i,j)|=p\}} [  \cR(\sigma(i,j),\tilde{\cF}_{k-1}^j) \\
&  \oplus \cR_{\text{N}}(\sigma(i,j), \B( \frac{\epsilon}{\max \{ \Gamma\rho,1 \}} ) )    ] \\
& \subseteq \bigcup\limits_{p=1}^{\theta_M} \bigcup\limits_{\{ i\in\cY: |\sigma(i,j)|=p\}} [  \cR(\sigma(i,j),\tilde{\cF}_{k-1}^j)  \\
&  \oplus \Gamma \rho^{|\sigma(i,j)|} \B( \frac{\epsilon}{\max \{ \Gamma\rho,1 \}} ) ] \\
&  \subseteq (\bigcup\limits_{p=1}^{\theta_M} \bigcup\limits_{\{ i\in\cY: |\sigma(i,j)|=p\}}   \cR(\sigma(i,j),\tilde{\cF}_{k-1}^j)) \\
& \oplus \Gamma \rho\B( \frac{\epsilon}{\max \{ \Gamma\rho,1 \}} )    
 \subseteq \tilde{\cF}_k^{j}\oplus \B(\epsilon)\subseteq \cS_m^j\oplus \B(\epsilon).
\end{align*}
Taking into account that by invariance of $\{ \tilde{\cD}_k^j \}_{j\in\cV}$ it holds that $\cS_m^j\subseteq \tilde{\cD}_k^j$ for all $j\in\cV$, the result follows. $\blacksquare$

\subsection{Proof of Theorem~\ref{theorem99}}
(i) For any edge $(i,j,{\sigma})\in\cE$, we have 
$A_{\sigma}\cS^i\oplus\cW_\sigma  = A_{\sigma}(\check{\cS}_m^i  \bigcup\limits_{ \{  s\in\cV: |\sigma(s,i)|\leq T-1 \} }   \cR(\sigma(s,i), \check{\cS}_m^s ) )\oplus \cW_{\sigma} 
 \subseteq (A_{\sigma}\check{\cS}_m^i\oplus \cW_\sigma)\bigcup\limits_{ \{  s\in\cV: |\sigma(s,i)|\leq T-2 \}} (A_\sigma \cR(\sigma(s,i),\check{\cS}_m^s)\oplus\cW_\sigma) \cup \check{\cS}_m^j   \subseteq \check{\cS}_m^j \bigcup\limits_{ \{  s\in\cV: |\sigma(s,j)|\leq T-1 \}}  \cR(\sigma(s,j),\check{\cS}_m^s) = \cS^j.
$
Thus, $\{ \cS^j\}_{j\in\cV}$ is invariant with respect to the System. Consequently, $\cS^j\supseteq \cS_m^j$, for all $j\in\cV$.
On the other hand, by construction of the $T$-product Lifted System we have $\cS_m^j\supseteq \check{\cS}_m^j$, for all $j\in\cV$.
Taking into account that for any sequence $\sigma(i,j)$,  $(i,j)\in\cV\times \cV$ it holds  $\cR(\sigma(i,j),\cS_m^i)\subseteq \cS_m^j$ we have that
$\cS^j\subseteq \cS_m^j \bigcup\limits_{ \{ s\in\cV: |\sigma(i,j)|\leq T-1 \} } \cR(\sigma(i,j),\cS_m^s)\subseteq \cS_m^j,$
for all $j\in\cV$. Thus, it holds necessarily that $\cS^j=\cS_m^j$, for all $j\in\cV$.
(ii) By construction,  $x(0)\in \cS_M^{z(0)}\subseteq \cX_{z(0)}$  implies $x(T)\in \cS_M^{z(T)}$  for any switching sequence $\sigma(z(0),z(T))$, $(x(\cdot),z(\cdot))$ being
trajectories of the System \eqref{eq_sys1}--\eqref{eq_sys3} subject to the constraints \eqref{eq_con1}, \eqref{eq_con3}.
Consequently, $\check{\cS}_M^j\supseteq \cS_M^j$, for all $j\in\cV$.
In addition, for any $j\in\cV$,
$\cS_M^j\subseteq \cS_M^j\bigcap\limits_{ \{  d\in\cV: |\sigma(j,d)|\leq T-1   \}}\cC(\sigma(j,d), {\cS}_M^d)\subseteq
\check{\cS}_M^j\bigcap\limits_{ \{  d\in\cV: |\sigma(j,d)|\leq T-1   \}}\cC(\sigma(j,d), \check{\cS}_M^d)=\cS^j.$
Moreover, for any edge $(i,j,\sigma)\in\cE$, we have
$C(\sigma,\cS^j) $ 

$ =\cC(\sigma,\check{\cS}^M_j)\bigcap\limits_{\{ s\in\cV: |\sigma(j,d)|\leq T-1\}} \cC( \{ \sigma, \sigma(j,d) \}, \check{\cS}_M^d  )
 \subseteq \cC(\sigma,\check{\cS}^M_j)   \bigcap\limits_{\{ s\in\cV: |\sigma(i,d)|\leq T-1\}} \cC( \sigma(j,d), \check{\cS}_M^d  )\cap \check{\cS}_M^i=\cS^i$, 
thus, $\{ \cS^j\}_{j\in\cV}$ is invariant and necessarily, $\cS^j \subseteq \cS_M^j$, $j\in\cV$.
Thus, $\cS^j=\cS_M^j$, for all $j\in\cV$.  $\blacksquare$

\subsection{Proof of Theorem~\ref{theorem11}}
(i)
Let $\{ \cF_l^j \}_{j\in\cV}$ and $\{\check{\cF}_l^j \}_{j\in\cV_P}$ denote the members of the multi-set sequences of the System and the $P$-path-dependent lifted system generated by \eqref{eq_setseq_min1_1},
\eqref{eq_setseq_min1_2}. We show that
\begin{equation} \label{eq_th11_3}
\cF_l^j=\cup_{i\in\cI(j)}\check{\cF}_l^i, \quad \quad j\in\cV.
\end{equation}
For $l=0$ we have $\cF_0^j=\cup_{i\in\cI(j)}\check{\cF}_0^i=\{ 0\}$. Assuming \eqref{eq_th11_3} holds for $l=k$, we have for $l=k+1$ 
$\cF_{k+1}^j  = \bigcup\limits_{(s_j,j,\sigma)\in\cE}\cR(\sigma,\cF_k^{s_j})= \bigcup\limits_{(s_j,j,\sigma)\in\cE} \cR(\sigma,  \bigcup\limits_{i\in\cI(s_j)}\check{\cF}_k^i	 ) 
 =  \bigcup\limits_{(s_j,j,\sigma)\in\cE} \bigcup\limits_{i\in\cI(s_j)} \cR\left(\sigma,  \check{\cF}_k^i	 \right) = \bigcup\limits_{i\in\cI(s_j)} \bigcup\limits_{(s_j,j,\sigma)\in\cE} \cR\left(\sigma,  \check{\cF}_k^i	 \right) =  \bigcup\limits_{i\in\cI(j)}\check{\cF}_{k+1}^i,
$
thus, \eqref{eq_th11_3} holds for all $l\geq 0$. Taking the limit as $l\rightarrow \infty$, the result follows.
(ii) It is straightforward to show that the members of the multi-set sequences $\{ \cB_l^j \}_{j\in\cV}$
and $\{\check{\cB}_{l}^j \}_{j\in\cV_P}$, generated by \eqref{eq_setseq_max1_1}, \eqref{eq_setseq_max1_2}
for the System and the $P$-Path-Dependent Lifted System satisfy $\cB_l^j=\check{\cB}_l^i$, for all $j\in\cV$, for all $i\in\cI(j)$, thus, the relation \eqref{eq_th12_1} follows by taking the limit as $l\rightarrow \infty$.  $\blacksquare$

\subsection{Proof of Proposition~\ref{theorem13}}
We first show the multi-set sequence \eqref{eq_th13_1} is a fixed point of an appropriately initialized backward reachability multi-set sequence \eqref{eq_setseq_max1_2}.
Consider $\{ \cS_0^j \}_{j\in\cV}$, with $\cS_0^j=\tilde{\cS}_M^j$, $j\in\cY$ and $\cS_0^j=\cX$, $j\in\cV\setminus \cY$.
For each unavoidable node $i\in\cY$ we have
$\cS_1^i=\tilde{\cS}_M^i\cap (  \bigcap\limits_{(i,d,\sigma_i)\in\cE}\cC(\sigma_i,\cS_0^d) )= \tilde{\cS}_M^i\cap  \cC(\sigma_i,\tilde{\cS}_M^i) \cap\cC(\sigma_i,\cX) = \tilde{\cS}_M^i$,
since $\cS_0^d=\cX$ and $\cC(\sigma_i,\cX)\supseteq \cC(\sigma_i,\tilde{\cS}_M^i)\supseteq \tilde{\cS}_M^i$.
Moreover, for each node $j\in\cV\setminus \cY$ in the path $m(i,d)$, $(i,d)\in\cY\times \cY$, for which $|\sigma(j,d)|=1$, it holds that
$\cS_1^j={\cX}\cap\cC(\sigma_i,\tilde{\cS}_M^d)$.
For any $j\in m(i,d)\setminus \{i\}$, $|\sigma(j,d)|>1$, it holds that $\cS_1^j=\cX\cap\cC(\sigma_i,\cX)$.
Next, for any $i\in \cY$ we have $\cS_2^i=\tilde{\cS}_M^i\cap (  \bigcap\limits_{(i,d,\sigma_j)\in\cE}\cC(\sigma_i,\cS_1^d) )=
\tilde{\cS}_M^i\cap \cC(\sigma_i,\cX\cap\cC(\sigma_i,\cX))=\tilde{\cS}_M^i\cap \cC(\sigma_i,\cX)\cap\cC(\{\sigma_i,\sigma_i\},\cX))=\tilde{\cS}_M^i$, since  $\cC(\{\sigma_i,...,\sigma_i\},\cX)\supseteq\cC(\{\sigma_i,...,\sigma_i\},\tilde{\cS}_M^i)=\tilde{\cS}_M^i$.
Moreover, for any $j\in m(i,d)$, $(i,d)\in\cY\times \cY$, for which $|\sigma(j,d)|=1$, it holds that $\cS_2^i=\cS_1^i$, while for all $j\in m(i,d)$, $(i,d)\in\cY\times \cY$, for which $2\leq|\sigma(j,d)|\leq\tau-1$
it holds that $\cS_2^j=\cX\cap\cC(\sigma_i,\cX)\cap \cC(\{\sigma_i,\sigma_i\},\tilde{\cS}_M^d)$.
By iterating $\tau-1$ times, the multi-set sequence $\{ \cS_{\tau-1}^j \}_{j\in\cV}$ is equal to \eqref{eq_th13_1}.
Additionally, we can verify that for each node $j\in\cV$, $\cS_{\tau}^j=\cS_0^j \bigcap \limits_{(j,d,\sigma)\in \cE} \cC(\sigma,\cS_{\tau-1}^d)=\cS_{\tau-1}^j$, thus, by Theorem~\ref{theorem3},
the multi-set $\{ \cS_\tau^j \}_{j\in\cV}$ is an admissible invariant multi-set with respect to the System \eqref{eq_sys1}-\eqref{eq_sys3} with constraints $\cX_j=\tilde{\cS}_M^j$, $j\in\cY$, $\cX_j=\cX$, $j\in\cV\setminus \cY$. By invariance of the multi-set $\{\cS_\tau^j \}_{j\in\cV}$ it holds that ${\cS}_\tau^j\subseteq \cS_M^j$, $j\in\cY$ and by taking into account that by construction $\cS_\tau^j=\tilde{\cS}_M^j\supseteq \cS_M^j$, $j\in\cY$, it  follows that $\tilde{\cS}_M^j=\cS_M^j$, for all $j\in\cY$. To show that $\cS_\tau^j=\cS_M^j$, for all $j\in\cV\setminus\cY$ we pick any path $m(i,d)$, $(i,d)\in\cY\times \cY$ and we consider the node $j$ for which
 $\sigma(j,d)=1$. Then, $\cS_M^{j}=\cX\cap\cC(\sigma_i,\cS_M^d)=\cS_{\tau}^j$. Continuing sequentially in the same manner until the node $j$ such that $|\sigma(i,j)|=1$, we have that $\cS_M^j=\cS_\tau^j$ for all $j\in m(i,d)$. Thus, the maximal invariant multi-set with respect to the System \eqref{eq_sys1}--\eqref{eq_sys3} is given by \eqref{eq_th10_1}.   $\blacksquare$

\subsection{Proof of Proposition~\ref{theorem14}}
As in Proposition~\ref{theorem13}, we can show that \eqref{eq_th14_1}
is a fixed point of the backward reachability multi-set sequence initialized by
$\cS_0^1=\tilde{\cS}_M^1$, $\cS_0^j=\cX$, $j\in\cV\setminus \{1\}$, which is retrieved in exactly $\tau$ steps.
Admissibility and invariance of the multi-set \eqref{eq_th14_1} follow from the initial multi-set $\{ \cS_0\}_{j\in\cV}$ and Theorem~\ref{theorem3} respectively.
To show maximality, we observe first for the unavoidable node that $\tilde{\cS}_M^1\subseteq \cS_M^1\subseteq \tilde{\cS}_M^1$. The result follows by applying similar steps as
in the last part of Theorem~\ref{theorem13}. $\blacksquare$

\subsection{Proof of Proposition~\ref{proposition7}}
Given any $T\geq 1$, we observe  the pair $(\Gamma_T, \rho_T)$ can be assigned as a stability metric for the T-product Lifted System, with $\Gamma_T:=\Gamma$, $\rho_T:=\rho^T$.  By applying Theorem~\ref{theorem3}, we can compute the
maximal invariant set $\check{\cS}_M$ of the T-product Lifted System in
$\check{k}=\left\lceil \log_{\rho_T}\left(\frac{R}{\Gamma c}\right)\right\rceil=
\left\lceil \frac{k}{T}\right\rceil$. Taking into account from \eqref{eq_th10_1} that
additional $T-1$ basic iterations are required for computing the maximal invariant set, the total number of iterations for computing the maximal invariant set in two steps is $g(T):= \left\lceil \frac{k}{T}   +T-1 \right\rceil$.
The optimal lift $T^\star$ that minimizes the function $g(\cdot)$
is $T^\star=\sqrt{k}$. The result \eqref{eq_proposition7_1} is reached by computing $g(T^\star)$. $\blacksquare$

\section{How to compute the pair $(\Gamma,\rho)$ in Example~\ref{example33} }
Under Assumption~\ref{ass3} and following the same reasoning as in \cite[Theorem 1]{Nikos14}, we pick\footnote{From \cite[Theorem 1]{Nikos14}, 
any subunitary choice of $\lambda_0$ is valid.} $\lambda_0=0.15\in (0,1)$ and compute the integer
 $k_0=96$ such that  $\s{\cN}^j_{k_0}\subseteq \lambda_0 \s{\cN}_0^j$, for all $j\in\cV$, where $\{\s{\cN}_{l}^j \}_{j\in\cV}$, $l\geq 0$, is provided by \eqref{eq_convnom1}, \eqref{eq_convnom2}. In this example we have $\s{\cN}_0^j=\cN_{0}^j$ which is a convex set. Consequently, by linearity of the dynamics, the pair  $(\Gamma,\rho)$ in \eqref{eq_lemma1_1} is retrieved by setting
 $ \Gamma   = \lambda_0^{\frac{-k_0+1}{k_0}} \max\limits_{j\in\cV}\min \{ \Gamma:        \conv{\bigcup\limits_{i\in[0,k_0-1]} \s{\cN}_i^j}\subseteq \Gamma \cN_0^j    \}=12.6023$,
 $ \rho  = \lambda_0^{\frac{1}{k_0}}=0.9804$.
Another way of computing the pair $(\Gamma,\rho)$ is by considering the Lyapunov theoretic framework in \cite{MatRaph:15}. To this purpose, by applying a $T$--product Lift, $T=4$, we obtain the multi-set $\{\cL^j \}_{j\in\cV}$ that satisfies the inclusions
$\cR_{\text{N}}(\sigma(i,j),\cL^i)\subseteq \gamma \cL^j$, $\gamma=0.9104$, for all $i\in\cV$ for which $|\sigma(i,j)|=T$, for all $j\in \cV$.
By computing the positive scalars $\alpha_1=1.5882, \alpha_2=3.5954$ which
$\alpha_1 \cL^j\subseteq \cN_0^j\subseteq \alpha_2\cL^j$, for all $j\in\cV$, it follows that $\cN_{tT}^j\subseteq \frac{\alpha_2}{\alpha_1}\gamma^t\cN_0^j$, for all $t\geq 0$.
By choosing the integer $t^\star=25$ we obtain the set inclusions $\cN_{k_0}^j\subseteq \lambda_0 \cN_0^j$, $j\in\cV$, with $\lambda_0:=\frac{\alpha_2}{\alpha_1}\gamma^{t^\star}=0.2166$, $k_0:=t^\star T=100$. Consequently, we obtain a second pair $(\Gamma',\rho')$ that satisfies \eqref{eq_lemma1_1}, with $\Gamma{'}=8.7666$, $\rho{'}=0.9848$.

\balance

\bibliographystyle{elsarticle-harv}
\bibliography{nikosb}
\end{document}